\documentclass[preprint]{aastex63}

\usepackage{url}
\usepackage{soul}

\newcommand{\ssst}{\scriptscriptstyle}
\newcommand{\E}[1]{\times 10^{#1}}

\newcommand{\s}{\,{\rm s}}      
\newcommand{\ps}{\,{\rm s}^{-1}}

\newcommand{\cm}{\,{\rm cm}}    
\newcommand{\km}{\,{\rm km}}
\newcommand{\kms}{$\km\ps$}

\newcommand{\kpc}{\,{\rm kpc}}

\newcommand{\K}{\,{\rm K}}

\newcommand{\pcc}{\,{\rm cm}^{-3}}
\newcommand{\um}{\,\mu\rm m}

\newcommand{\Tmb}{T_{\rm mb}}

\newcommand{\VLSR}{V_{\ssst\rm LSR}}
       
\newcommand{\gray}{{\rm $\gamma$-ray}}
\newcommand{\grays}{{\rm $\gamma$-rays}}

\newcommand{\Fermi}{{\sl Fermi}}

\newcommand{\snr}{G35.6$-$0.4}
\newcommand{\seast}{G35.6$-$0.5}

\newcommand{\twCO}{$^{12}$CO}   
\newcommand{\thCO}{$^{13}$CO}

\newcommand{\Jotz}{$J$=1--0}

\submitjournal{ApJ}

\shorttitle{Mult-wavelength study of SNR G35.6$-$0.4}
\shortauthors{Zhang et al.}

\graphicspath{{./}{img/}}

\begin{document}
\title{GeV Gamma-ray Emission and Molecular Clouds towards Supernova Remnant G35.6$-$0.4 and the TeV Source HESS J1858+020}

\correspondingauthor{Xiao Zhang \& Yang Chen}
\email{xiaozhang@nju.edu.cn; ygchen@nju.edu.cn}

\author[0000-0002-9392-547X]{Xiao Zhang}
\affiliation{School of Astronomy \& Space Science, Nanjing University, 163 Xianlin Avenue, Nanjing~210023, China}

\affiliation{Key Laboratory of Modern Astronomy and Astrophysics, Nanjing University, Ministry of Education, Nanjing~210023, China}

\author[0000-0002-4753-2798]{Yang Chen}
\affiliation{School of Astronomy \& Space Science, Nanjing University, 163 Xianlin Avenue, Nanjing~210023, China}
\affiliation{Key Laboratory of Modern Astronomy and Astrophysics, Nanjing University, Ministry of Education, Nanjing~210023, China}

\author{Fa-xiang, Zheng}
\affiliation{School of Astronomy \& Space Science, Nanjing University, 163 Xianlin Avenue, Nanjing~210023, China}

\author[0000-0002-5786-7268]{Qian-Cheng, Liu}
\affiliation{School of Astronomy \& Space Science, Nanjing University, 163 Xianlin Avenue, Nanjing~210023, China}

\author[0000-0002-5683-822X]{Ping Zhou}
\affiliation{School of Astronomy \& Space Science, Nanjing University, 163 Xianlin Avenue, Nanjing~210023, China}
\affiliation{Key Laboratory of Modern Astronomy and Astrophysics, Nanjing University, Ministry of Education, Nanjing~210023, China}

\author{Bing, Liu}
\affiliation{Department of Astronomy, School of Physical Sciences, University of Science and Technology of China, Hefei, Anhui 230026, China}
\affiliation{CAS Key Laboratory for Research in Galaxies and Cosmology, University of Science and Technology of China, Hefei, Anhui 230026, China}
\affiliation{Key Laboratory of Modern Astronomy and Astrophysics, Nanjing University, Ministry of Education, Nanjing~210023, China}

\begin{abstract}

It is difficult to distinguish hadronic process from the leptonic one in \gray\ observation, which is however crucial in revealing the origin of cosmic rays.
As an endeavor in the regard,
we focus in this work on the complex \gray\ emitting region, which partially overlaps with the unidentified TeV source HESS~J1858+020 and includes supernova remnant (SNR) G35.6$-$0.4 and HII region G35.6$-$0.5.
We reanalyze CO-line, HI, and \Fermi-LAT GeV \gray\ emission data of this region. 
The analysis of the molecular and HI data suggests that SNR G35.6$-$0.4 and HII region G35.6$-$0.5 are located at different distances. 
The analysis the GeV \grays\ shows that GeV emission arises from two point sources: one (SrcA) coincident with the SNR, and the other (SrcB) coincident with both HESS J1858+020 and HII region G35.6$-$0.5. The GeV emission of SrcA can be explained by the hadronic process in the SNR-MC association scenario. The GeV-band spectrum of SrcB and the TeV-band spectrum of HESS J1858+020 can be smoothly connected by a power-law function, with an index of $\sim$2.2. The connected spectrum is well explained with a hadronic emission, with the cutoff energy of protons above 1 PeV. 
It thus indicates that there is a potential PeVatron in the HII region and should be further verified with ultra-high energy observations with, e.g., LHAASO.

\end{abstract}

\keywords{
supernova remnants --- 
ISM: individual (\snr) --- 
gamma rays: observations --- 
radiation mechanisms: nonthermal
}


\section{Introduction} \label{sec:intro}

In \gray\ astrophysics, multiband observations play an important role in identifying the source types and understanding the origin of the \gray\ emissions.
Thanks to the observations facilitated in the wide electromagnetic wave window, about 2/3 of the TeV-\gray\ sources have counterparts at other wavelengths.
Because of this, it was konwn that there are a number of types of Galactic \gray\ emitters, such as supernova remnants (SNRs), pulsars (PSRs) and their wind nebulae (PWNe), HII regions (HIIRs; or star-forming regions), superbubbles, etc.
These Galactic \gray\ sources may be the accelerators of cosmic ray (CRs) below the ``knee" energy around $\sim3\times10^{15}$ eV if they have the hadronic origin and have the maximum energy of \gray{s} above $\sim$100 TeV.
For example, based on the observations toward the typical HIIRs in Milky way, e.g., the Cygnus Region \citep{veritas2014.cygnus,argo2014.cygnus,asgamma2021.cygnus,hawc2021.cygnus}, it is suggested that HIIRs are efficient particle accelerators and may be a type of contributors of Galactic CRs.
Due to the TeV spectrum without an obvious cutoff, the unidentified TeV source HESS J1858+020 \citep{aharonian08,hess2018.hgps}, likely associated with an SNR-HIIR complex\footnote{The term ``complex" here does not necessarily mean that the SNR and the HIIR are physically related.} G35.6$-$0.4, may be a good candidate of CR source and provide an unique opportunity to distinguish the contribution of high energy \grays\ from the two components.

\begin{figure}
\centering
\includegraphics[height=82mm,angle=0]{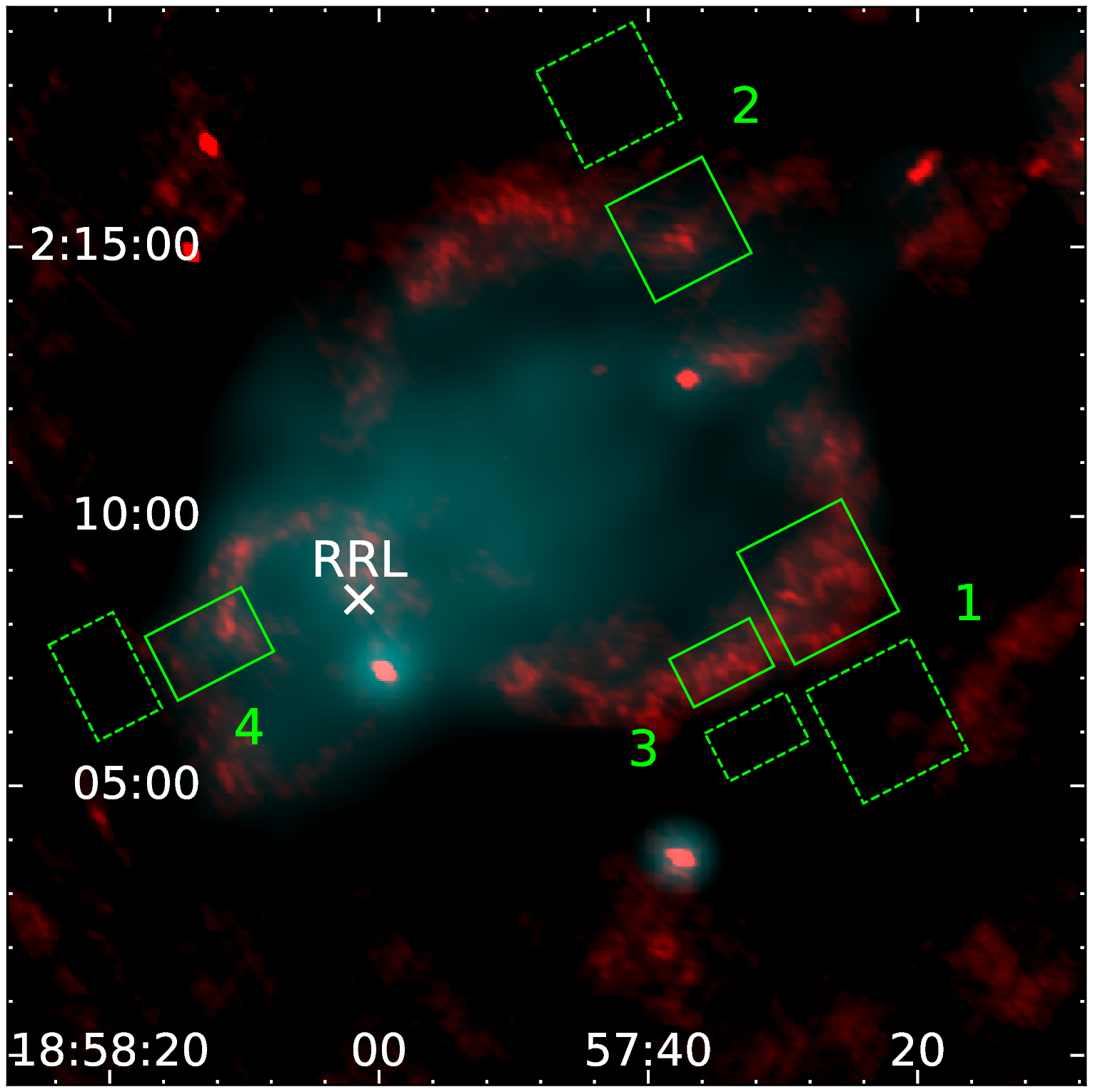}
\includegraphics[height=82mm,angle=0]{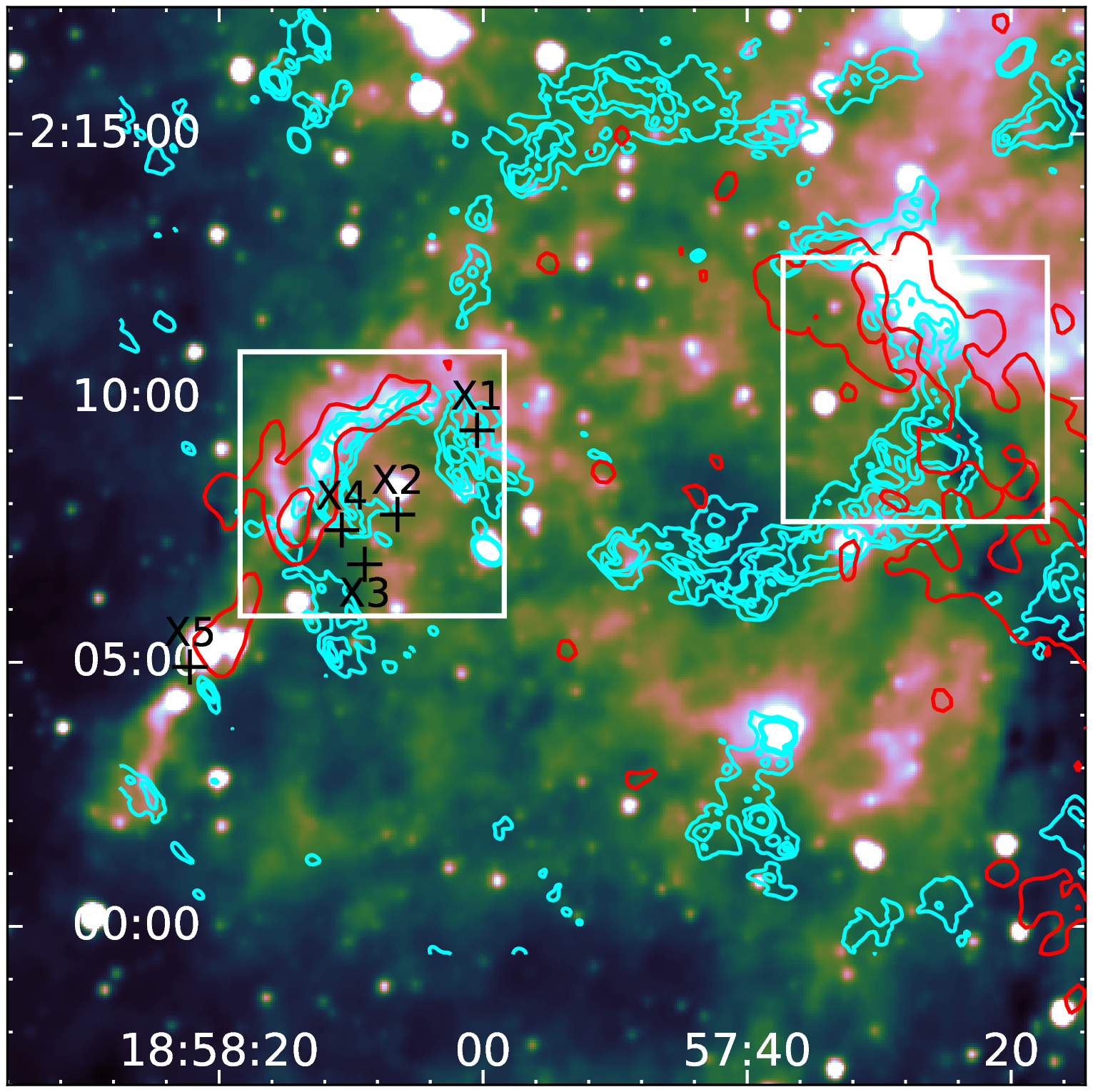}
\caption{
Colored images of the SNR-HIIR complex \snr\ in equatorial coordinate system. Left: The 610 MHz and 1.4 GHz data are represented in red and cyan, respectively. The green solid and dashed boxes mark the source and background regions for studying the HI absorption, respectively. The white cross represents the RRLs found by \citet{lockman89}. Right: WISE 12$\um$ image superposed with GMRT 610 MHz contours (cyan) in levels of 0.44, 0.66, and 0.88$\,{\rm mJy\,beam^{-1}}$ \citep{paredes14} and Nobeyama $^{12}$CO~$(J=1-0)$ contours (red) in levels of 15 and 30$\,{\rm K}\,\km\ps$ in the velocity range of $+53 - +57\km\ps$. 
The two white boxes represent the regions used to plot the grids of CO spectra. The black crosses are the X-ray sources found by \citet{paredes14}.
}
\label{fig:rgb}
\end{figure}

The extended radio source G35.6$-$0.4 was discovered in the early radio surveys \citep{beard69,altenhoff70} and had a longtime debate on its class: an HIIR, an SNR, or an SNR-HIIR complex \citep[see][]{green09}.
According to the non-thermal spectral index $\alpha_r=0.47\pm0.07\ (S\propto \nu^{-\alpha_r}$, where $S$ is the radio flux and $\nu$ the frequency) and the faint infrared (IR) emission, it was re-identified as an SNR although radio recombination lines are detected from it \citep{green09}.
Its radio morphology revealed by the VGPS 1.4 GHz data \citep{stil06,green09} shows a partially limb-brightened structure elongated in the Galactic latitude with a size of $15'\times11'$. 
With deep observations performed by the Giant Metrewave Radio Telescope (GMRT), however, it was resolved at 610 MHz into two nearly circular sources which appear connected with each other \citep{paredes14}.
This high-resolution GMRT image suggests that the radio source G35.6$-$0.4 defined in the previous works consists of an SNR (G35.6$-$0.4) represented by the big circular structure and an HIIR (G35.6$-$0.5), containing the radio recombination lines \citep[RRLs,][]{lockman89}, represented by the ring-shaped structure with a smaller size (see the left panel of Figure~\ref{fig:rgb}).
Moreover, there is a semi-circular structure in WISE 12$\um$ image \citep{Wright2010} which well delineates the  small radio semi-ring (see the right panel of Figure~\ref{fig:rgb}).

The distance to the SNR was first assumed to be the same as that, $\sim$10.5 kpc, to the nearby HIIR G35.5$-$0.0, which seems consistent with the $\Sigma-D$ relation \citep{green09}.
It was also put at a small distance of $3.6\pm 0.4$ kpc based on the HI absorption feature abstracted from the radio continuum brightness peak region \citep{zhu13} and the association with a CO cloud at the local-standard-of-rest (LSR) velocity $\VLSR\sim+55\ {\rm km\ s^{-1}}$ as suggested by \citet{paron10}. 
The small distance was also supported by the latest HI study \citep{ranasinghe18}.
As will be shown below, however, SNR G35.6$-$0.4 and HIIR  G35.6$-$0.5 are located at different distances and are not in physical association with one another.

At very high-energy band, a TeV source HESS J1858+020 was found to partially overlap with the southern (in the Galactic coordinate system) part of SNR \snr\ \citep{aharonian08}.
Assuming a purely hadronic origin, a lower limit on the proton cutoff energy is derived as $\sim$30 TeV at 90\% confidence level \citep{Spengler2020}.
Based on the CO observation, in the overlapped region there is a  molecular cloud (MC) which was suggested to be likely associated with the SNR and the TeV $\gamma$-ray emission detected in HESS J1858+020 was ascribed to SNR-MC hadronic interaction \citep{paron10,paron11}.
At GeV band, no emission was found from the 2-yr {\it Fermi}-LAT data \citep{torres11}.
Recently, \citet{cui21.g35.6} performed a {\sl Fermi}-LAT \gray\ analysis with 10.7-yr data and found that there are two GeV point sources toward the \snr{} region from the spatial study in the energy range 5--500 GeV.
One hard source (SrcX2) is spatially coincident with HESS J1858+020 and a molecular clump.
It was suggested that the GeV-TeV \gray\ emission is from the MC bombarded by the protons escaped from the SNR \citep{cui21.g35.6}.
The other (SrcX1) with a soft GeV spectrum is located at the northern boundary of the SNR, with the origin of the \grays\  remaining unclear.

In this work, we revisit the {\it Fermi}-LAT observational data of the \gray\ emitting sources toward the SNR-HIIR complex G35.6$-$0.4 and perform a study of the  molecular environment of the complex. 
Data used in this work and the corresponding results are given in sections~\ref{sec:data} and \ref{sec:results}, respectively.
In section~\ref{sec:discussion}, the revised distances of the two emitting sources and the origin of the \gray\ emissions are discussed.
Finally, we summarize our results in section~\ref{sec:summ}.

\section{Observations and Data}
\label{sec:data}

\subsection{\Fermi-LAT Observational data}

In this study, we analyze more than 12 years (from 2008-08-04 15:43:36 (UTC) to 2020-11-30 05:38:25 (UTC)) of {\sl Fermi}-LAT Pass 8 SOURCE class (evclass=128, evtype=3) data in the energy range 0.2-500~GeV using 
the python package \emph{Fermipy} (v1.0.1)\footnote{\url{https://fermipy.readthedocs.io/en/latest/}} \citep{fermipy}.
The corresponding instrument respond functions (IRFs) are ``P8R3\_SOURCE\_V3\_v1".
The region of interest (ROI) is a $15\arcdeg\times 15\arcdeg$ square centered at the position of SNR \snr\ (RA = 284.479$\arcdeg$, Dec = 2.217$\arcdeg$).
We only select the events within a maximum zenith angle of 90$\degr$ to filter out the background \grays\ from the Earth's limb and apply the recommended filter string ``($\rm DATA\_QUAL>0) \&\& (LAT\_CONFIG==1$)" to choose the good time intervals. The Galactic diffuse emission (\emph{gll\_iem\_v07.fits}) and isotropic emission (\emph{iso\_P8R3\_SOURCE\_V3\_v1.txt}), 
as well as all the sources listed in the LAT 10-year Source Catalog  \citep[4FGL-DR2,][]{abdollahi20} within a radius of $25\degr$ from the ROI center, are included for background modeling, namely the baseline model.

\subsection{CO observations and archival data}

The observations of \twCO~\Jotz~(at 115.271 GHz) and \thCO~\Jotz~(at 110.201 GHz) were made during 2013 April-May with the 13.7-m millimeter-wavelength telescope of the Purple Mountain Observatory at Delingha (PMOD), China.
The total bandwidth of the fast Fourier transform spectrometer of PMOD is 1 GHz and the half-power beamwidth is about 50$''$ for the two lines.
The typical RMS noise level is about 0.5 K for \twCO~(\Jotz) at the velocity resolution of 0.16~\kms\ and 0.3 K for \thCO~(\Jotz) at 0.17~\kms. 

We also use the archival \twCO\ data of the FOREST Unbiased Galactic plane Imaging survey with the Nobeyama 45-m telescope \citep[FUGIN,][]{umemoto17} observation. The data have an angular resolution of $20''$ and an average RMS of 1.5~K at a velocity resolution of 0.65~\kms.

\subsection{Other data}

We also retrieve the Very Large Array (VLA) radio continuum image at 1.4 GHz and HI data from the HI/OH/recombination line survey (THOR) project \citep{beuther16,wang20}. The data have a spatial resolution of $25''$ and a velocity resolution of 1.5 \kms.

\section{Data analysis and Results}
\label{sec:results}
\subsection{\Fermi-LAT Data Analysis}

\subsubsection{Spatial Analysis}

\begin{figure}
\centering
\includegraphics[height=72mm,angle=0]{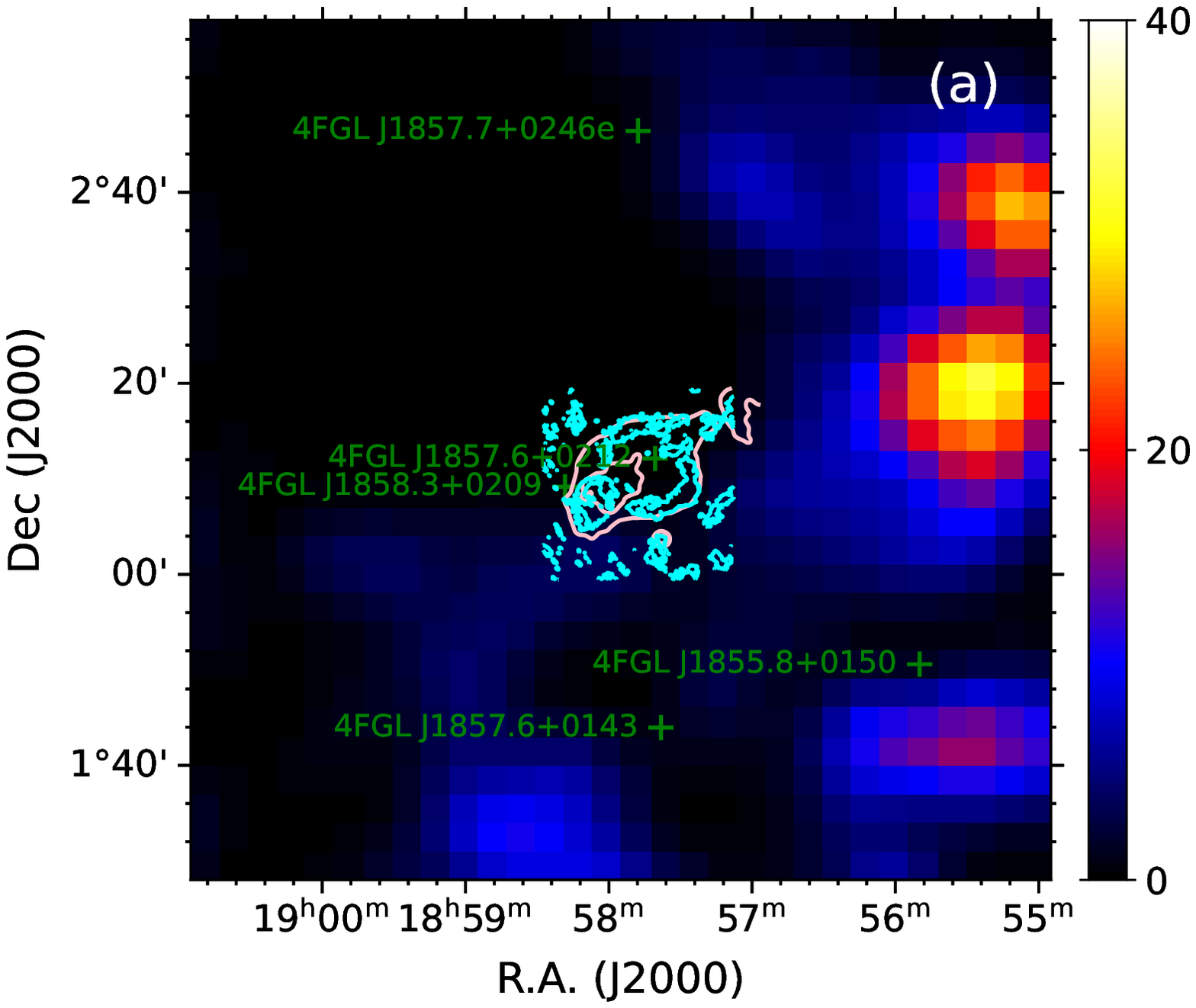}
\includegraphics[height=72mm,angle=0]{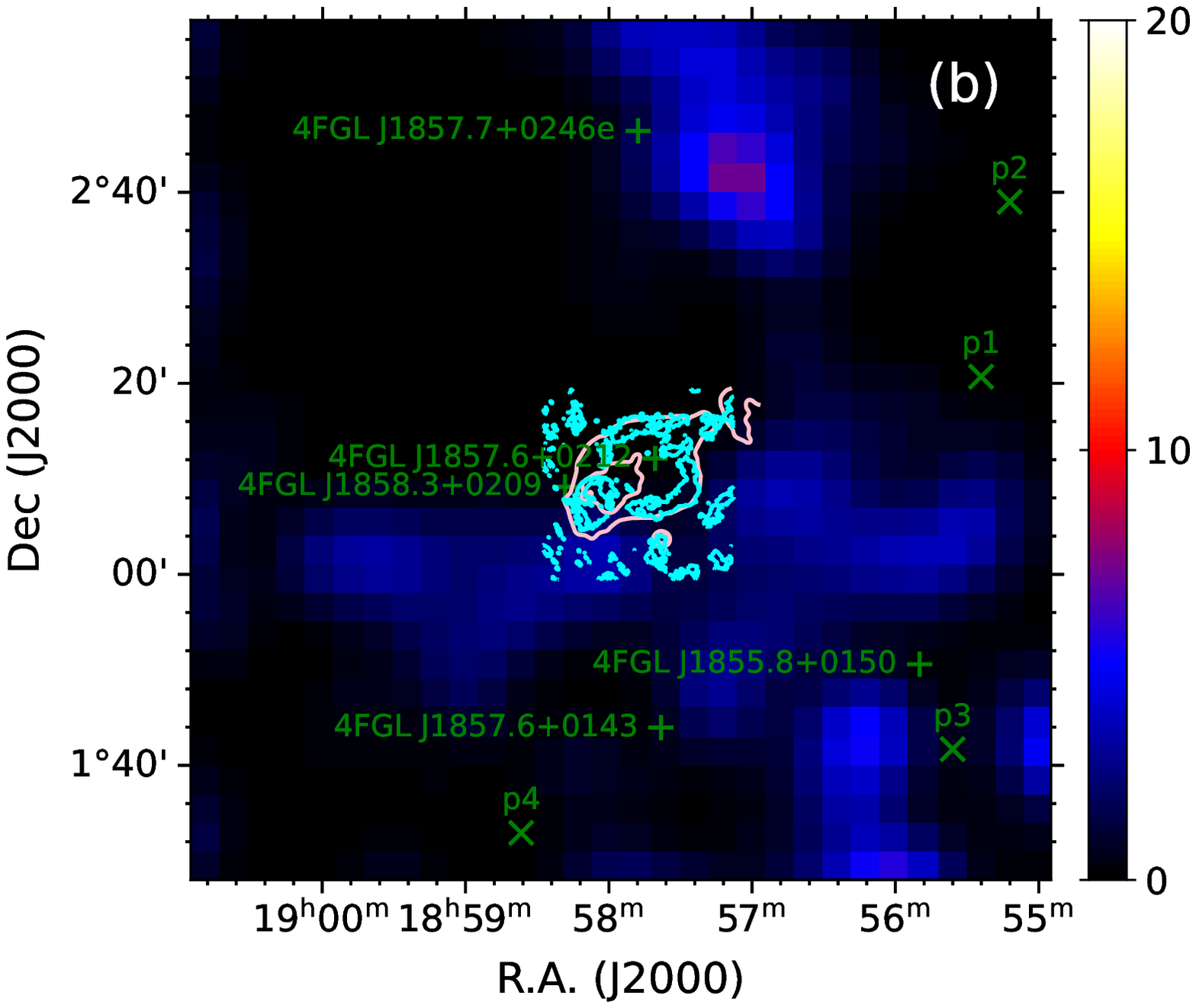}
\caption{
TS maps of a $1.5\degr\times1.5^{\circ}$ region centered at SNR \snr\ in the energy range of 0.2--500 GeV. Image is smoothed by the pixel size of $0.05^{\circ}$. The pink and cyan contours represent the SNR-HIIR complex \snr\ in 1.4 GHz \citep{stil06} and 610 MHz \citep{paredes14}, respectively. Left: baseline model; Right: four point sources p1--p4 with positions listed in Table~\ref{tab:addp} are included in the source model.
}
\label{fig:tsmap_p12}
\end{figure}

\begin{figure}
\centering
\includegraphics[height=72mm,angle=0]{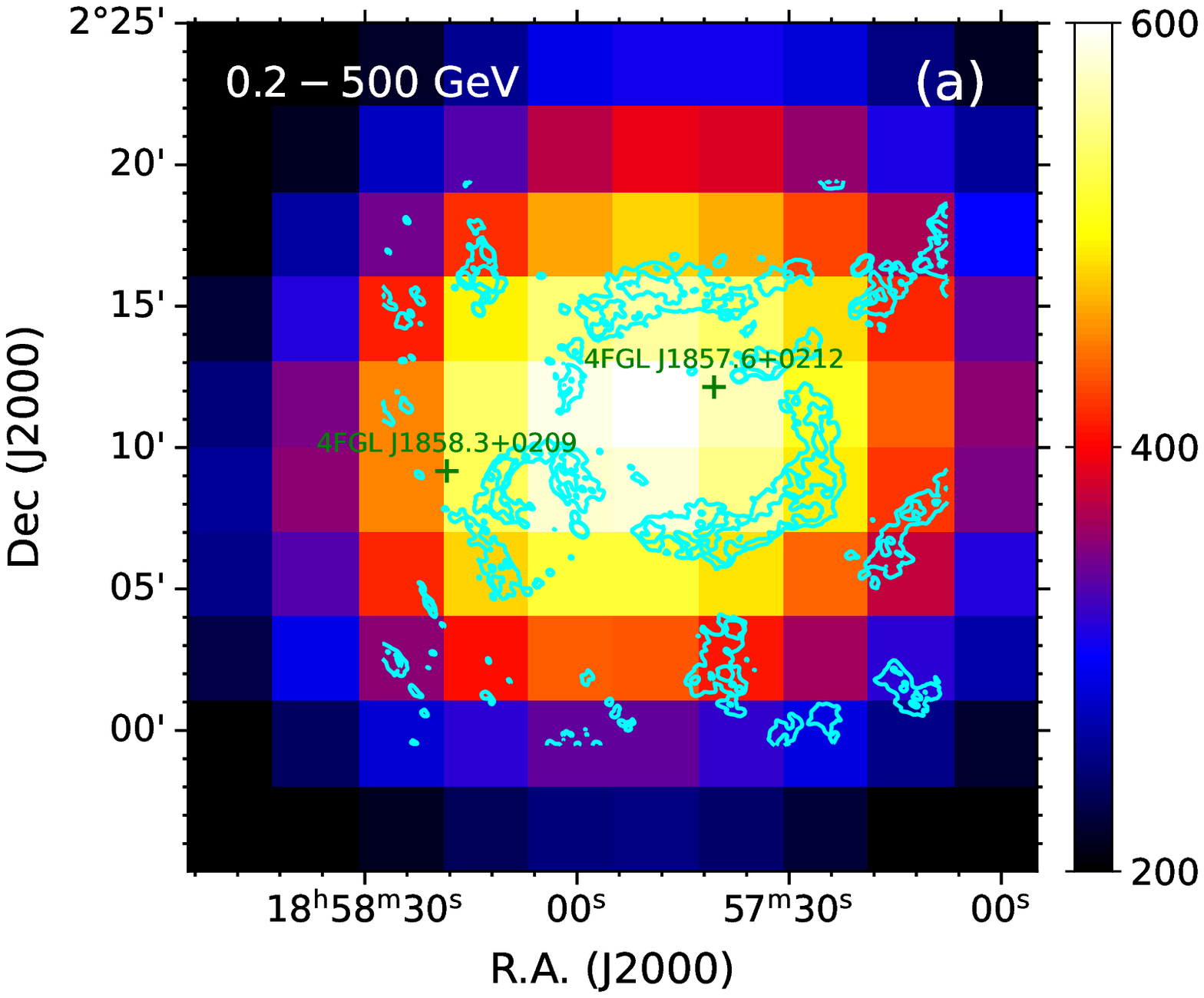}
\includegraphics[height=72mm,angle=0]{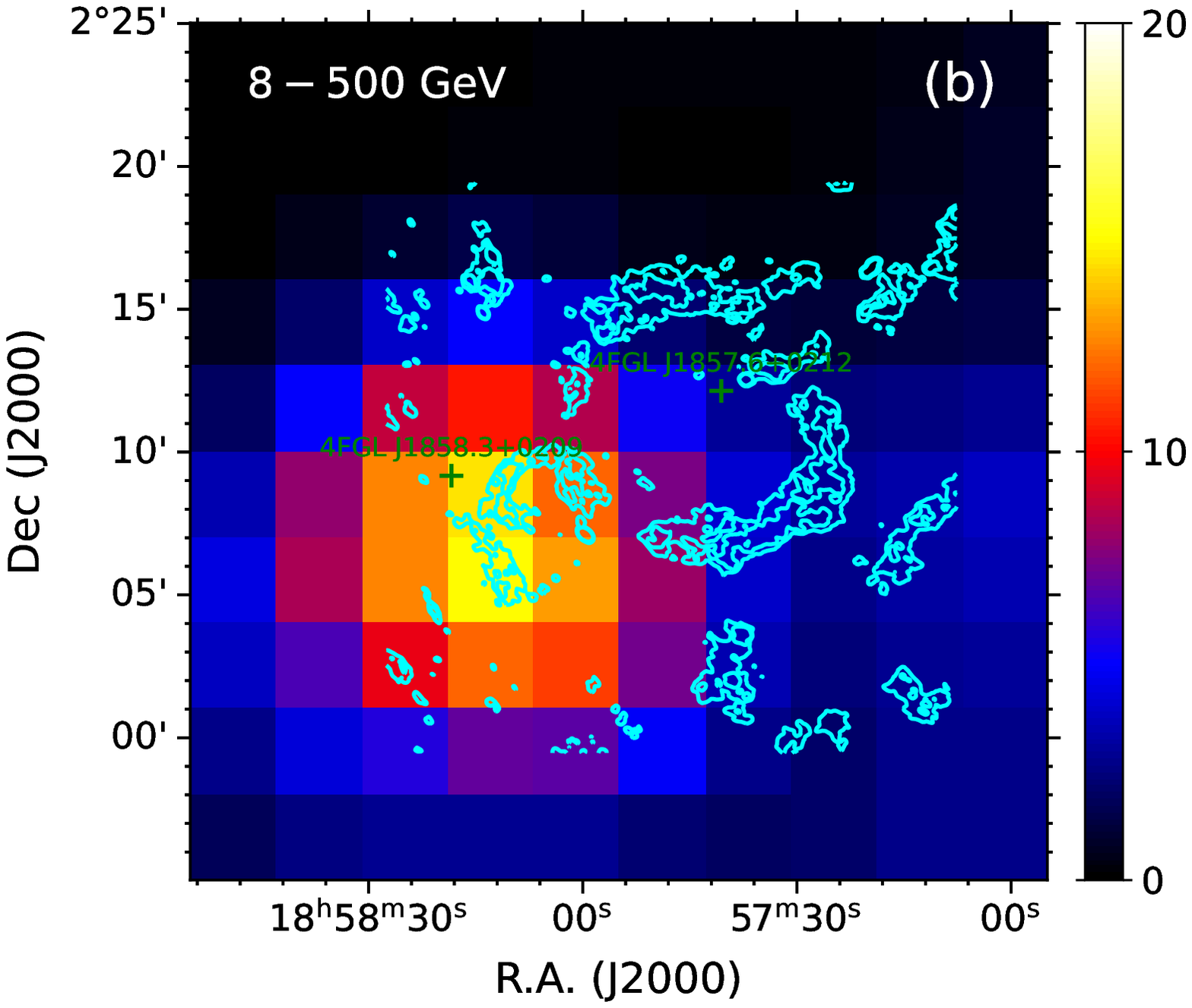}
\includegraphics[height=72mm,angle=0]{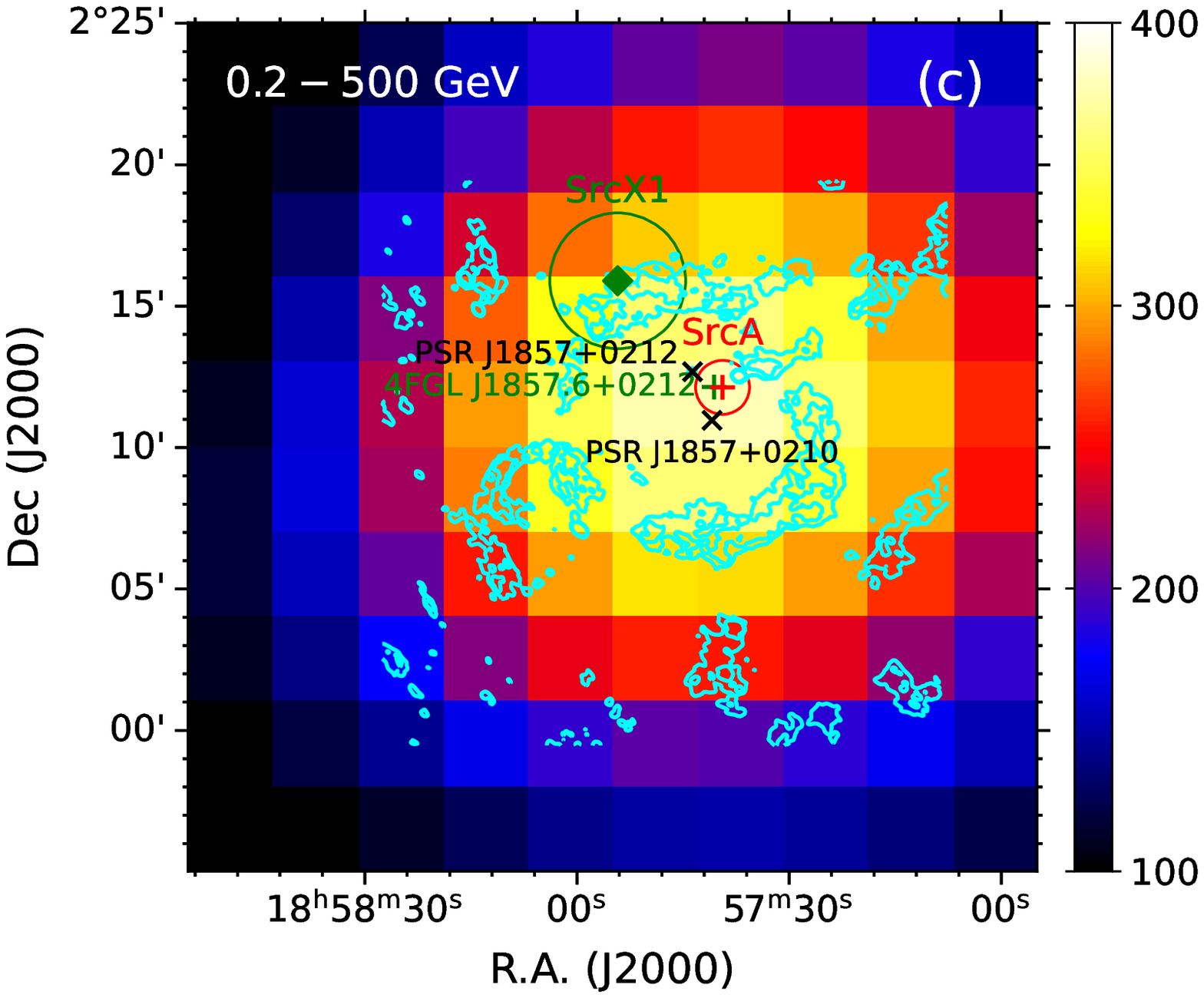}
\includegraphics[height=72mm,angle=0]{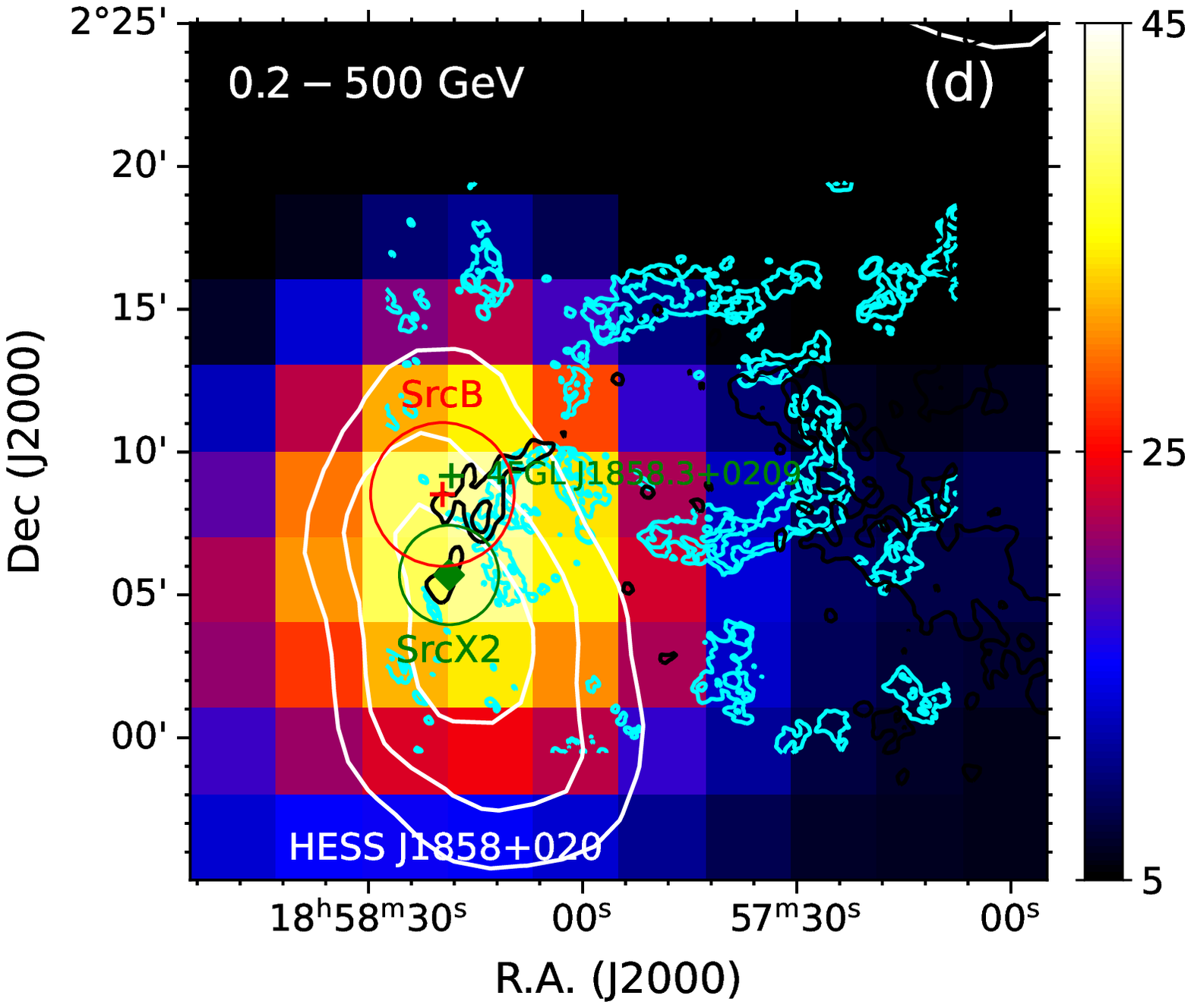}
\caption{
Same as Figure~\ref{fig:tsmap_p12} but in a $0.5\degr\times0.5^{\circ}$ region.
{\bf (a)} 4FGL J1857.6+0212 and 4FGL J1858.3+0209 are excluded from the source model in the energy range 0.2--500 GeV.
{\bf (b)} Same as (a) but in 8--500 GeV.
{\bf (c)} SrcA (4FGL J1857.6+0212) is excluded from the source model. Two pulsars are also marked with the black crosses.
{\bf (d)} Same as (c) but for SrcB (4FGL J1858.3+0209). The white and black contours show the TeV source HESS J1858+020 \citep{hess2018.hgps} with 5, 6, and 7$\sigma$ significance and the Nobeyama $^{12}\mathrm{CO}$  ($J$=1$-$0) emission with levels [15, 30]~K\,\kms\ in the velocity range of +53 -- +57 \kms, respectively.
For (c) and (d), the green diamonds show the best-fit positions obtained by \citet{cui21.g35.6} with 1$\sigma$ error shown in black circles. The refitted positions with 1$\sigma$ uncertainty in this study are marked by the red plus and circle. The original positions listed in 4FGL-DR2 are displayed in the green plus.
}
\label{fig:tsmap_ab}
\end{figure}

We note that there are two 4FGL-DR2 catalog sources 4FGL J1857.6+0212 and 4FGL J1858.3+0209 toward the SNR-HIIR complex \snr\ region. 
They are first treated as backgrounds to check the residual emission around our target.
We use {\sl fit} method to refit the spectral parameters of the sources within $3^{\circ}$ from the ROI center with the significance above $4\sigma$ and the normalization parameters of the two diffuse background components.
Then, we fix all parameters except for the normalization of the Galactic diffuse emission to their best-fit values and generate the residual test-statistic (TS) map which is shown in the left panel of Figure~\ref{fig:tsmap_p12}.
Here, the TS value is defined as TS = 2log($L_1/L_0$), in which $L_0$ is the maximum likelihood of the null hypothesis and $L_1$ the maximum likelihood with a putative source located in this pixel.
As can be seen in Figure~\ref{fig:tsmap_p12}a, there is still strong residual emission to the west of the SNR-HIIR complex. 
We thus add three point sources with simple powerlaw spectra located at the peak pixels with TS $>$ 20, the positions of which are listed in Table~\ref{tab:addp}, to model the residuals and repeat the above procedures.
After this, we find that there is still some excess to the south of the complex.
We then try adding this access as the fourth point source with powerlaw spectrum.
Compared to the 3-point-source model, the likelihood of 4-point-source model can be increased by about 15, resulting in $\mathrm{TS_{model}}$=2log$(L_{\rm 4ps}/L_{\rm 3ps})\approx30$.
So we use the 4 point sources to model the residual emission.
Finally, the background-subtracted TS map is displayed in Figure~\ref{fig:tsmap_p12}b. 

Next, we study the \gray\ morphology toward SNR-HIIR complex \snr\ in detail.
After excluding the 4FGL-DR2 catalog sources 4FGL J1857.6+0212 and 4FGL J1858.3+0209 from the source model, the TS maps in 0.2--500 GeV and 8--500 GeV are shown in Figures~\ref{fig:tsmap_ab}a and \ref{fig:tsmap_ab}b, respectively.
As can be seen, there is little emission above 8 GeV in the SNR region (represented by the bigger radio contours), while the emission at such energies concentrates on the HII region (represented by the smaller radio contours), suggesting that there are likely two sources.
Meanwhile, we use the likelihood ratio to test three spatial models: one point source, one uniform disk, and two point sources.
The model with the largest likelihood value will be preferred.
To do so, we first use one point source with a LogParabola (LogP) spectrum, and then explore its best-fit position and extension to obtain the likelihood values of $L_{\rm 1ps}$ for the one-point-source model and $L_{\rm disk}$ for the uniform-disk model.
For the two-point-source model, we use the templates of 4FGL J1857.6+0212 and 4FGL J1858.3+0209 and refit their best-fit positions, resulting in SrcA (RA = $284.418\degr$, Dec = $2.192\degr$, 1$\sigma$ error = $0.016\degr$) and SrcB (RA = $284.582\degr$, Dec = $2.142\degr$, 1$\sigma$ error = $0.042\degr$) for 4FGL J1857.6+0212 and 4FGL J1858.3+0209, respectively.
The extension of the two sources are also explored, giving ${\rm TS}_{\rm ext}=2.4$ and 4.0 for SrcA and SrcB, respectively.
Finally, according to the likelihood values for the three spatial models, we obtain $\mathrm{TS}_{\rm disk}=2\mathrm{log}(L_{\rm disk}/L_{\rm 1ps})=13$ and ${\rm TS}_{\rm 2ps}=30$ for the uniform-disk and two-point-source models, respectively.
Considering the TS distribution in the different energy ranges and the likelihood ratio test for the three spatial models, two-point-source model for the SNR-HIIR complex \snr\ is preferred and used in the following analysis.
The TS values of SrcA and SrcB in the two-point-source model is fitted to be 761.2 and 55.3, corresponding to the significance of 27.1$\sigma$ and 6.7$\sigma$, respectively.

In Figure~\ref{fig:tsmap_ab}, the TS maps for SrcA and SrcB are also presented in the second row panels.
Our best-fit positions with 1$\sigma$ uncertainty are displayed with the red pluses and circles, respectively.
For comparison, the 4FGL-DR2 positions and the best-fit ones in \citet{cui21.g35.6} are shown with green pluses and green diamonds, respectively.
As can be seen, the new best-fit positions for the both sources are very close to their original ones listed in the catalog, but are different from those in \citet{cui21.g35.6}, in particular for SrcA.
To check this difference, we fit the positions by just using 5--500 GeV data and obtain similar results to those of our above treatment in 0.2--500 GeV, ruling out the effect of different energy range used in the two studies.
Further, we repeat the analysis in \citet{cui21.g35.6} by replacing the IRFs ``P8R3\_SOURCE\_V3\_v1" with ``P8R3\_SOURCE\_V2\_v1" and using 8-yr catalog for the background, and reproduce the results in \citet{cui21.g35.6}.
Thus, the discrepancy in the best-fit position for SrcA is verified to be caused by using different IRFs and catalogs.


\begin{table}[h!]
\centering
\caption{Locations and TS value of the peak pixels in the left panel of Figure~\ref{fig:tsmap_p12}} \label{tab:addp}
\begin{tabular}{cccc}
\tablewidth{0pt}
\hline
\hline
 Name & R.A.(J2000) & Dec (J2000)  & TS \\
\hline
p1 & 283.850 & 2.345 & 41.5    \\
p2 & 283.800 & 2.650 & 39.1    \\
p3 & 283.900 & 1.695 & 23.5    \\
p4 & 284.653 & 1.549 & 16.5    \\
\hline
\end{tabular}
\end{table}

\subsubsection{Spectral Analysis}
In the 4FGL-DR2 catalog, the spectral types of SrcA and SrcB are LogP and PowerLaw (PL), respectively.
To study the spectral properties of the both sources in the whole energy range of 0.2--500 GeV, other spectral types including ExpCutoffPowerLaw (ECPL) and BrokenPowerLaw (BPL) will also be explored.
We first change the spectral type of SrcA to PL, ECPL, and BPL, respectively, to find the best spectral type.
At the same time, the spectral type of SrcB remains to be PL. 
After this, we keep the best choice for SrcA and change the spectral type of SrcB to LogP, ECPL, and BPL, respectively, to find the best spectral formula for SrcB. 
The formulae of these spectra are listed in Table~\ref{tab:spec}. 
The spectral type is favored if it has the largest TS value defined as $\mathrm{TS_{model}}=-2\mathrm{log}(L_\mathrm{PL}/L_\mathrm{model})$. 
As shown in Table~\ref{tab:ts}, a LogP spectrum is preferred for SrcA. 
But for SrcB, there is no obvious difference for the four spectral types and the simplest PL form is adopted.
Finally, we obtain the fluxes in 0.2--500 GeV as $4.2\times10^{-11}\ {\rm erg\ \cm^{-2}\ s^{-1}}$ (with $\Gamma=2.66\pm0.07$ and $\beta=0.44\pm0.06$ for $E_0$=1.57 GeV) and $1.1\times10^{-11}\ {\rm erg\ \cm^{-2}\  s^{-1}}$ (with $\Gamma=2.19\pm0.09$) for SrcA and SrcB, respectively.

The spectral energy distributions (SEDs) within 0.2--500\,GeV of SrcA and SrcB are generated by using the maximum likelihood analysis in seven logarithmically spaced energy bins.
During the fitting process, the free parameters only include the normalization parameters of the sources with the significance $4\sigma$ within $5^{\circ}$ from the ROI center as well as the Galactic and isotropic diffuse background components, while all the other parameters are fixed to their best-fit values from the above analysis in the whole energy (0.2--500 GeV) range.
In the energy bins where the TS value of SrcA or SrcB is smaller than 4, we calculate the 95\%-confidence-level upper limit of its flux.
The results are displayed in Figure~\ref{fig:sed}.

\begin{table}[h!]
\centering
\caption{Formulae for $\gamma$-ray spectra} \label{tab:spec}
\begin{tabular}{cll}
\tablewidth{0pt}
\hline
\hline
Name & Formula & Free parameters\\
\hline
\decimals
PL   & $dN/dE=N_0(E/E_0)^{-\Gamma}$     & $N_0, \Gamma$    \\
ECPL & $dN/dE=N_0(E/E_0)^{-\Gamma}\mathrm{exp}(-E/E_\mathrm{cut})$   &  $N_0,\Gamma,E_\mathrm{cut}$     \\
LogP & $dN/dE=N_0(E/E_0)^{-\Gamma-\beta\mathrm{log}(E/E_0)}$   &   $N_0,\Gamma,\beta$   \\
BPL  & $dN/dE=N_0 \begin{cases}{ (E/E_\mathrm{b})^{-\Gamma_1}, E\le E_\mathrm{b}\\ (E/E_\mathrm{b})^{-\Gamma_2}, E\ge E_\mathrm{b} }  \end{cases} $ &   $N_0, E_\mathrm{b}, \Gamma_1, \Gamma_2$   \\
\hline
\end{tabular}
\end{table}

\begin{table}[h!]
\centering
\caption{The Likelihood Test Results ($\mathrm{TS_{model}}$) from the Spectral Analysis of SrcA and SrcB} \label{tab:ts}
\begin{tabular}{ccccc}
\tablewidth{0pt}
\hline
\hline
   & $\mathrm{TS_{PL}}$ & $\mathrm{TS_{ECPL}}$ & $\mathrm{TS_{LogP}}$ & $\mathrm{TS_{BPL}}$\\
\hline
SrcA & 0 & 70.1 & 88.9 & 76.7    \\
SrcB & 0 & 0.1  & 1.7 & 2.3     \\
\hline
\end{tabular}
\end{table}

\subsection{Molecular environments}

\begin{figure}
\epsscale{1.1}
\plotone{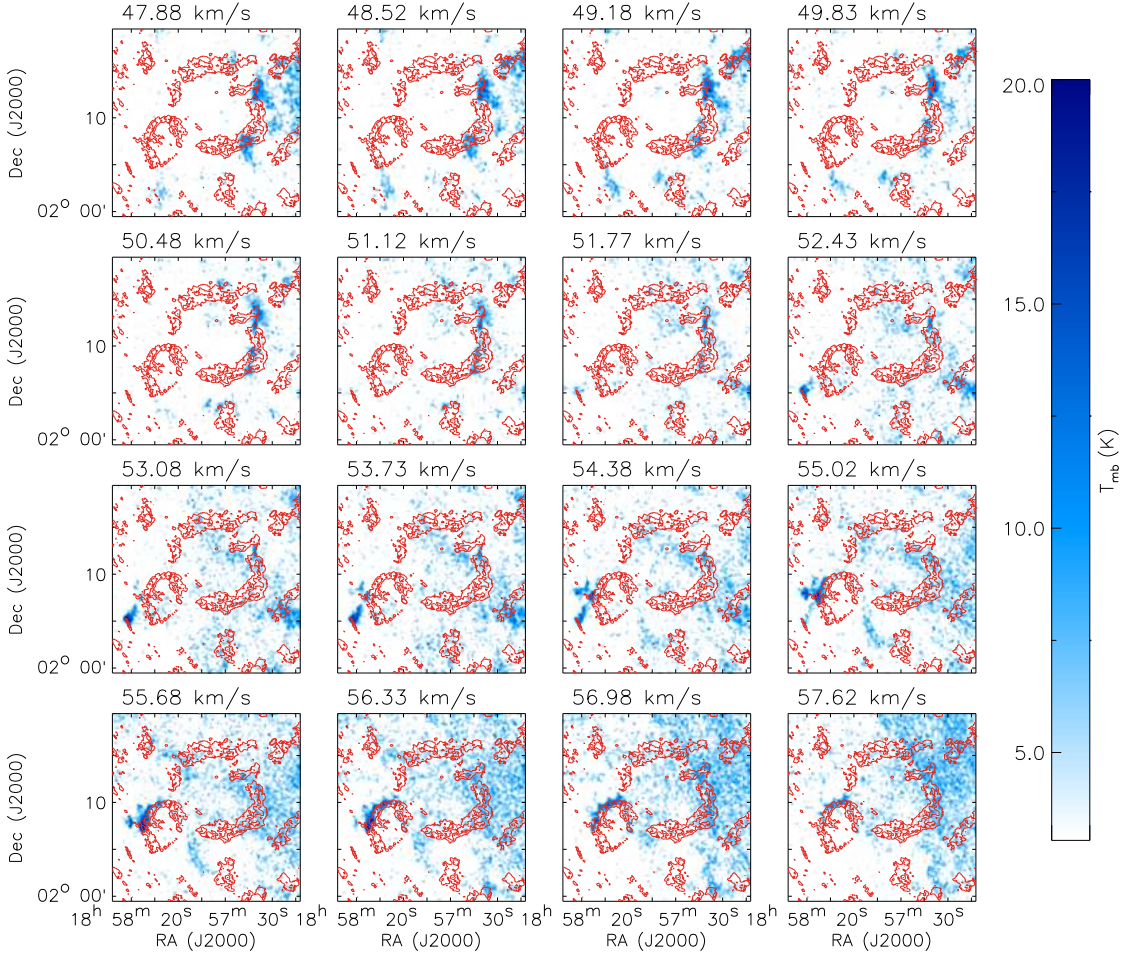}
\caption{
Channel map of the main-beam temperature of Nobeyama \twCO~\Jotz\ emission with a step of 0.65~\kms, overlaid with contours of GMRT 610~MHz radio emission. 
}
\label{fig:12co10_nob_grid}
\end{figure}

The \twCO\ spectra toward SNR \snr\ show multiple line components from $+6\km\ps$ to $+100\km\ps$. At
$\VLSR$ lower than $\sim +40\km\ps$ or higher than $\sim +70\km\ps$, little 
spatial correspondence between the MCs and the SNR can be
found from the CO emission.
Figure~\ref{fig:12co10_nob_grid} shows the spatial distribution of the Nobeyama \twCO\ emission in velocity range $\sim +47$ to $+58\km\ps$, with velocity intervals of $0.65\km\ps$.
The molecular gas in this velocity range seems to generally have a spatial correspondence with the western edge of SNR \snr.
In particular, a bright, clearcut molecular filament nicely follows the western shell of SNR \snr\ at $\VLSR\sim +50\km\ps$.
Notably, the peak \twCO\ main-beam temperature of the filament is about 20\,K, which is noticeably higher than the typical temperature of $\sim 10$\,K in interstellar MCs. 

\begin{figure}
\epsscale{1.1}
\plotone{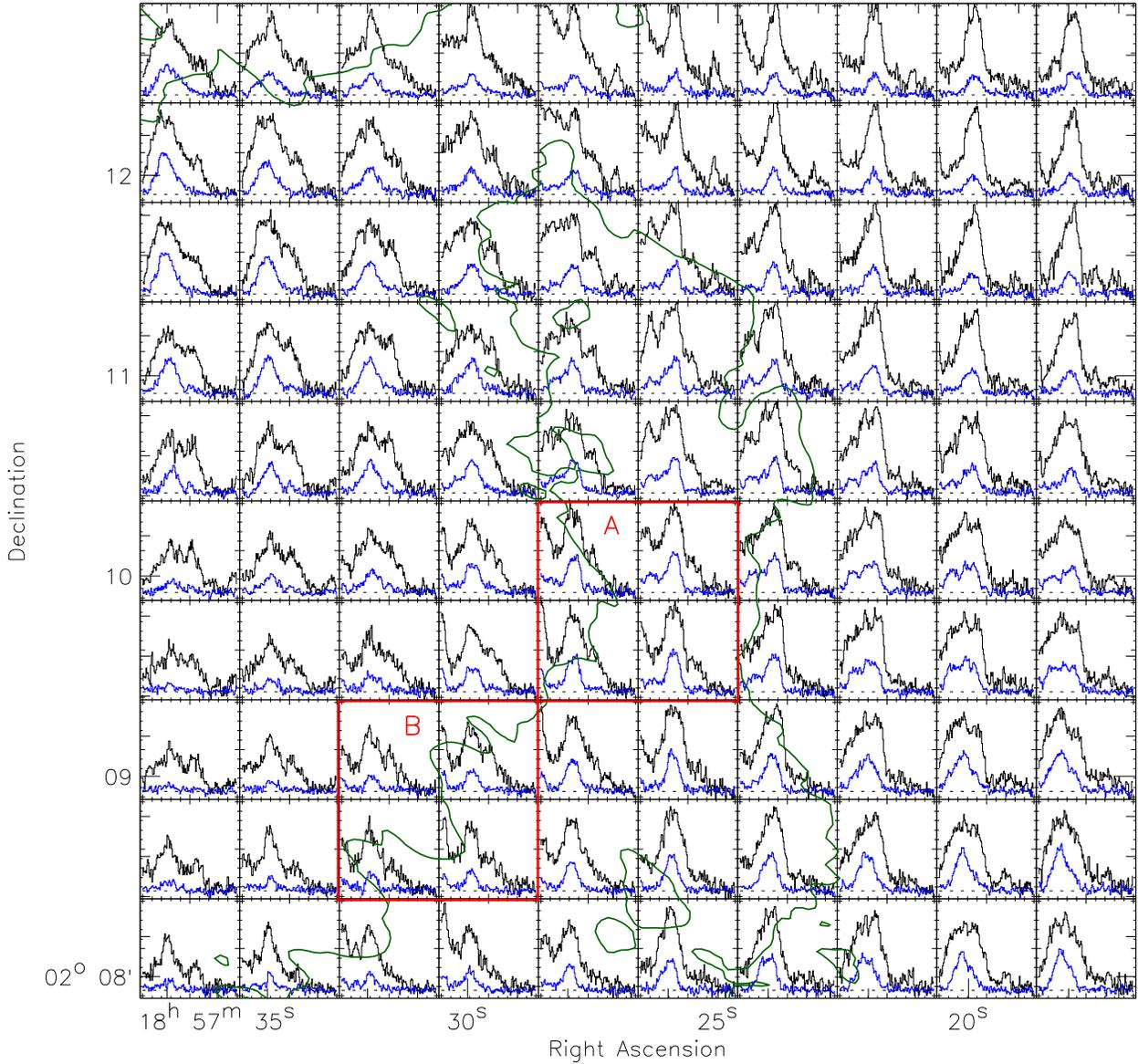}
\caption{
Grid of PMOD \twCO\ (black) and \thCO\ (blue) line profiles in the velocity range of $+50$ -- $+70$~\kms for the region delineated by the right white box marked in Figure~\ref{fig:rgb} in the SNR region.
The GMRT 610 MHz contour of 0.44 mJy beam$^{-1}$ is shown in green.
The size of each pixel is $30''\times 30''$. 
The averaged spectra from regions ``A'' and ``B'' are shown in Figure~\ref{fig:co_spec}.
}
\label{fig:co_linegrid_w}
\end{figure}

\begin{figure}
\epsscale{0.5}
\plotone{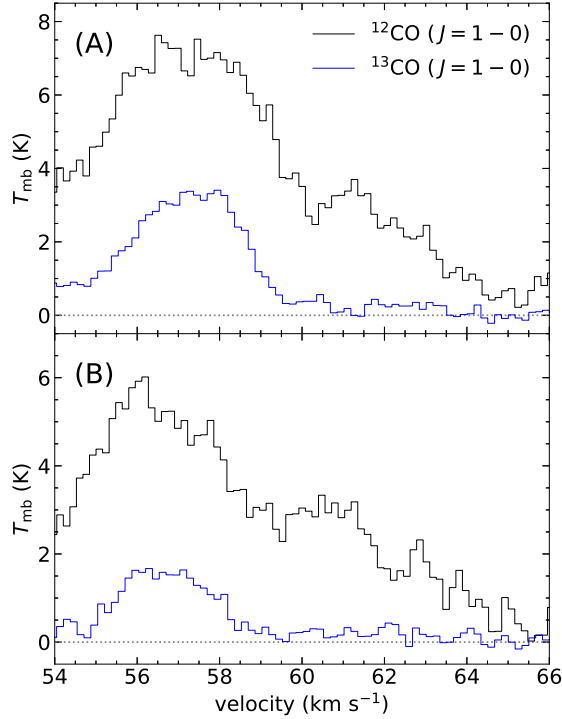}
\caption{
PMOD CO spectra in the $+54$ -- $+66$~\kms interval for the two regions marked in Figure~\ref{fig:co_linegrid_w}. The black and blue lines are for the \twCO\,($J=1-0$) and \thCO\,($J=1-0$), respectively.
}
\label{fig:co_spec}
\end{figure}

To investigate whether the molecular gas at the western edge of SNR \snr\ in the velocity range is perturbed by the shock of the SNR shock, we inspect the CO line grid toward this region.
As shown in a grid of both \twCO\ (\Jotz) and \thCO (\Jotz) spectra covering the western edge of the SNR at $\VLSR \sim +50$ -- $+70\km\ps$ (see Figure~\ref{fig:co_linegrid_w}),
the red wings of the \twCO\ lines, which peak at around $+57\km\ps$, seem to be asymmetrically broadened from about $+59$ to $+64\km\ps$, and these features are only presented within the boundary of the SNR.
Figure~\ref{fig:co_spec} shows the averaged CO
line profiles of a few pixels at the western edge of the SNR (regions `A' and `B', as marked in Figure~\ref{fig:co_linegrid_w}). 
In these regions, at $\VLSR\sim +59$ --  $+64\km\ps$, there are non-negligible \twCO\ intensities while there is little  \thCO\ emission. 
Because the \thCO\ is usually optically thin and yielded in quiescent, intrinsically high-column-density molecular material, an asymmetrical \twCO-line profile without a similar \thCO\ counterpart could be a kinematic signature of shock perturbation of the molecular gas.
Therefore, the broadened \twCO\ red wings indicate that the MCs in this velocity range are probably perturbed by SNR \snr.

\begin{figure}
\epsscale{1.1}
\plotone{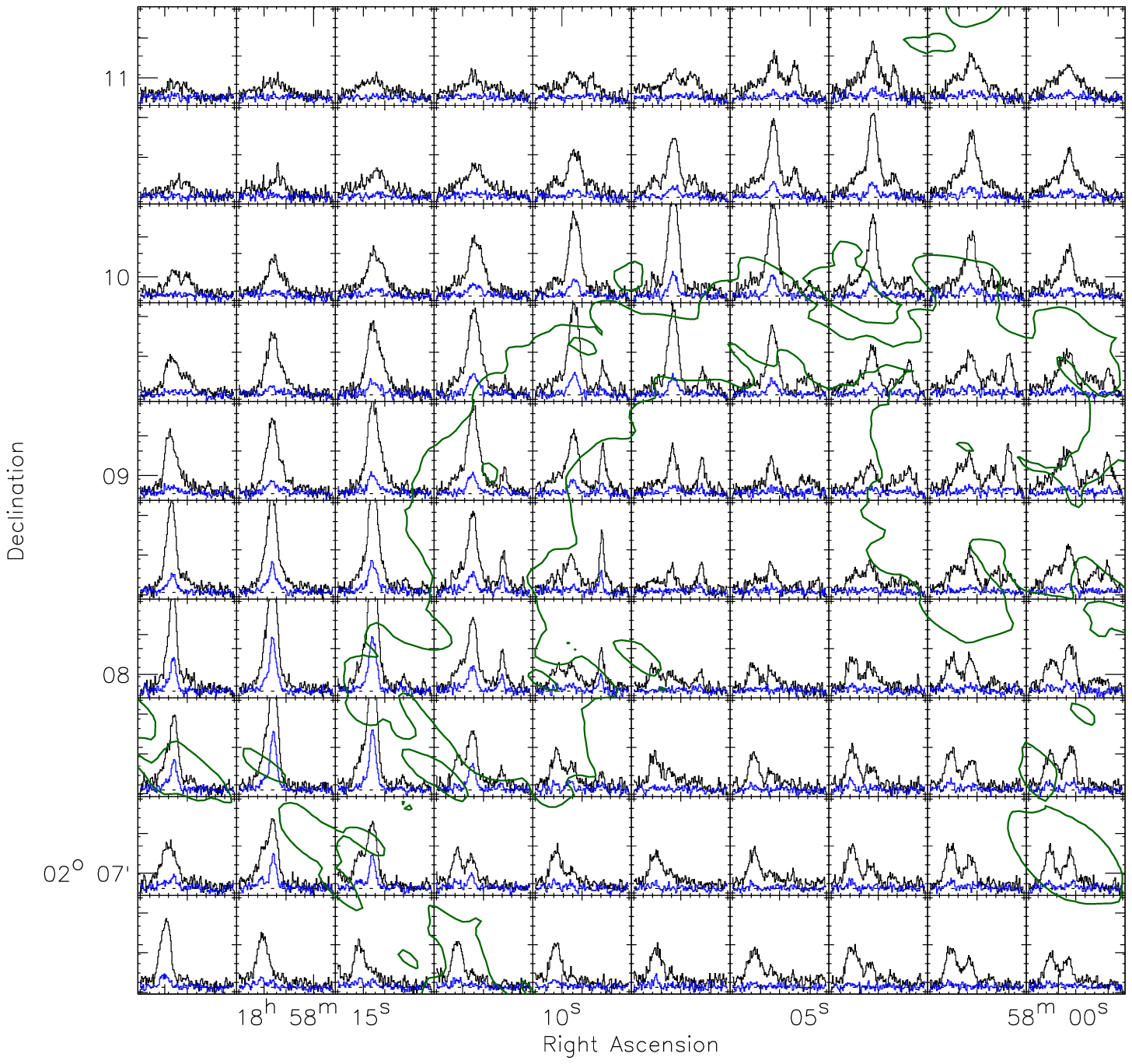}
\caption{
Same as Figure~\ref{fig:co_linegrid_w} but for HIIR in the velocity range of $+48$ -- $+68$~\kms.
}
\label{fig:co_linegrid_e}
\end{figure}

Toward HIIR \seast, there some noteworthy features at similar LSR velocities.
First, there is a thin \twCO\ molecular arc at $\VLSR\sim +55$--$+57~\km\ps$, which corresponds to the ``northern clump" described in \citet{paron10} and delineates the semi-ring of the HIIR, consistent with the previous \thCO\ studies \citep{paron10,paron11}.
The peak main-beam temperature at the molecular arc is $\Tmb\sim 22~\K$, which is also noticeably higher than the typical temperature of $\sim 10$~K in quiescent interstellar MCs. 
Secondly,
the wings of both the \twCO\ (\Jotz) and the \thCO\ (\Jotz) lines at a few positions along the molecular arc seem to be slightly asymmetric (see Figure~\ref{fig:co_linegrid_e}).
Actually, the asymmetry of the \thCO\ lines have been noted in \citet{paron10}.
The morphological agreement, together with the relatively high temperature and the asymmetric CO line profiles, supports the association between the molecular arc and the HII region.
Thirdly, a cloudlet at $+52$ -- $+55\km\ps$, corresponding to the ``southern" (in the Galactic coordinated system) clump in \citet{paron10}, is present outside the southeastern edge of the HIIR (see Figure~\ref{fig:12co10_nob_grid}) and is spatially close to the centroid of HESS J1858+020 (see Figure~\ref{fig:tsmap_ab}d). 
Especially, the peak of the main-beam temperature of the cloudlet ($\ga 20$\,K) is also significantly high.

\section{Discussion}
\label{sec:discussion}

\subsection{Distances to the HIIR and SNR}


Based on the association with the +55 -- +56\kms\ MCs suggested by \citet{paron10}, the distance to SNR-HIIR complex \snr\ was constrained as $3.6\pm0.4$ kpc by the HI absorption feature \citep{zhu13}.
The HI absorption spectra abstracted from the radio
continuum brightness peak region (very close to the position of RRLs) show the maximum absorption velocity around +61 \kms, supporting the near distance.
However, the radio image of SNR \snr\ presented in previous works \citep[e.g.,][]{green09} was later resolved into two circular structures by GMRT \citep{paredes14}.
Although these two structures are projectively connected with each other, they may be at different distances.
Thus the HI absorption spectra in \citet{zhu13} abstracted from the overlapped region of the two circles maybe only give constraint on the distance to one of them.
To separately give constraint on the distances to the two structures, we examine HI absorption by selecting other regions marked in green boxes in Figure\,\ref{fig:rgb} and avoiding the overlapped region of the two circles.

\begin{figure}
  \centering
\includegraphics[angle=0, width=0.45\textwidth]{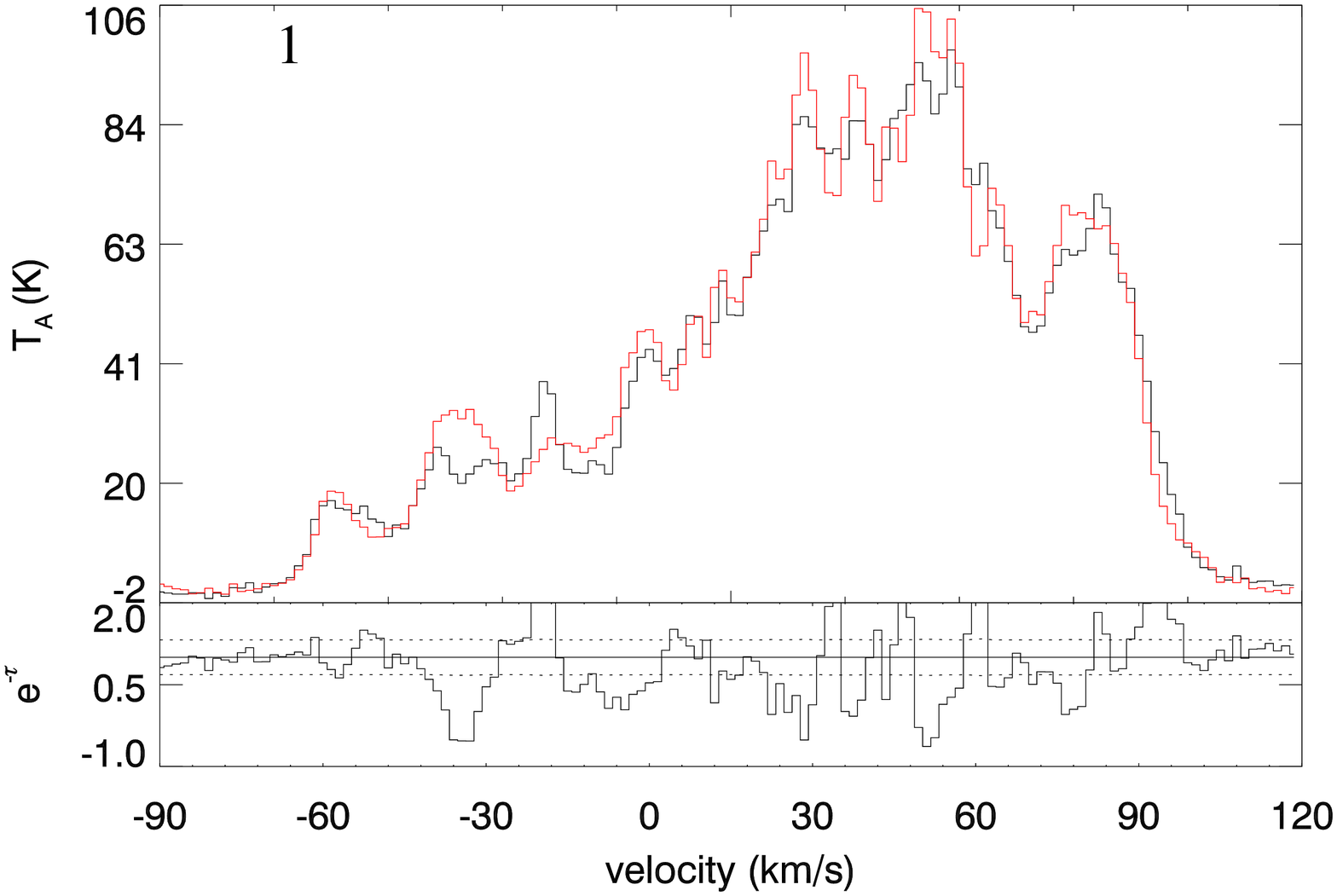}
\includegraphics[angle=0, width=0.45\textwidth]{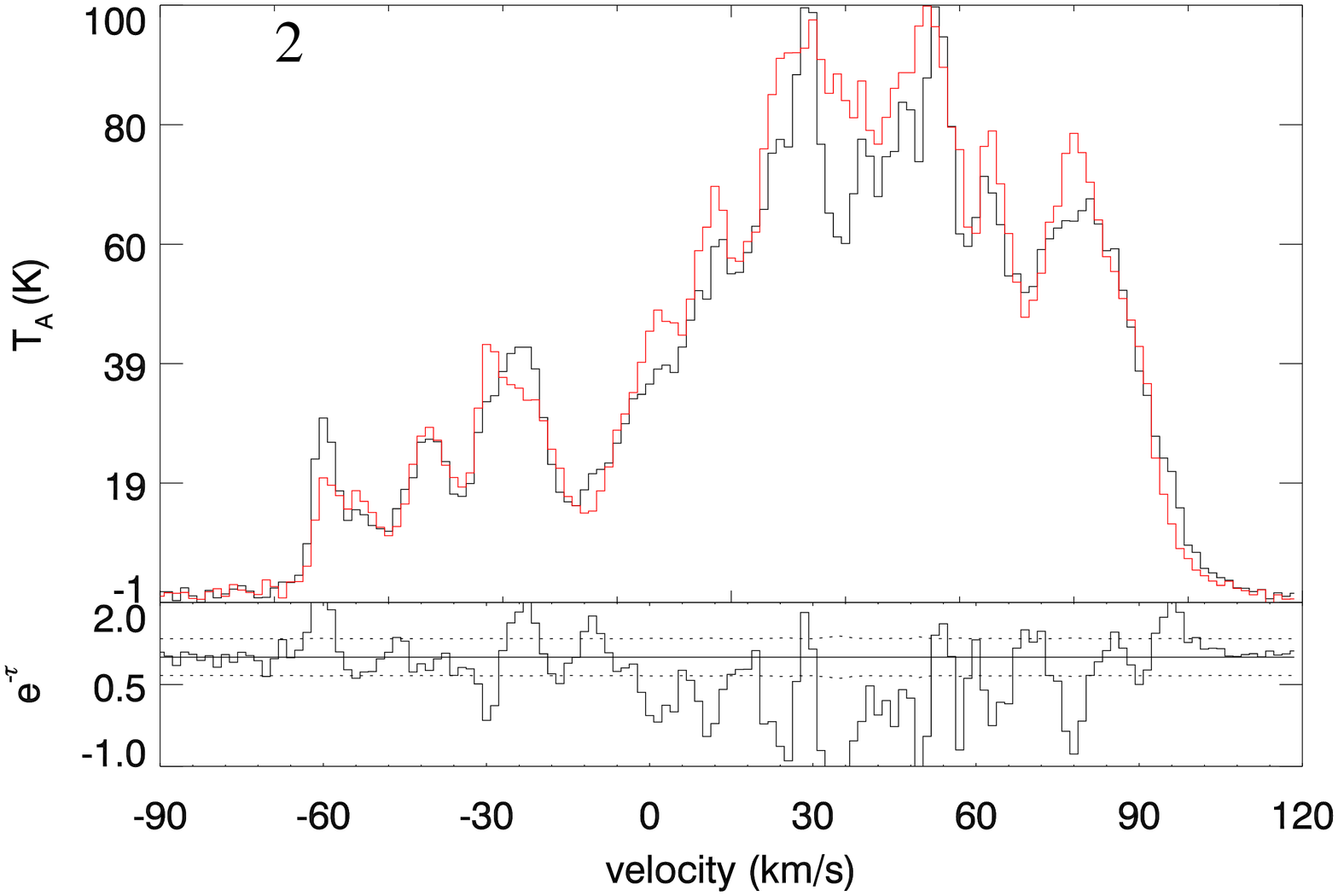}
\includegraphics[angle=0, width=0.45\textwidth]{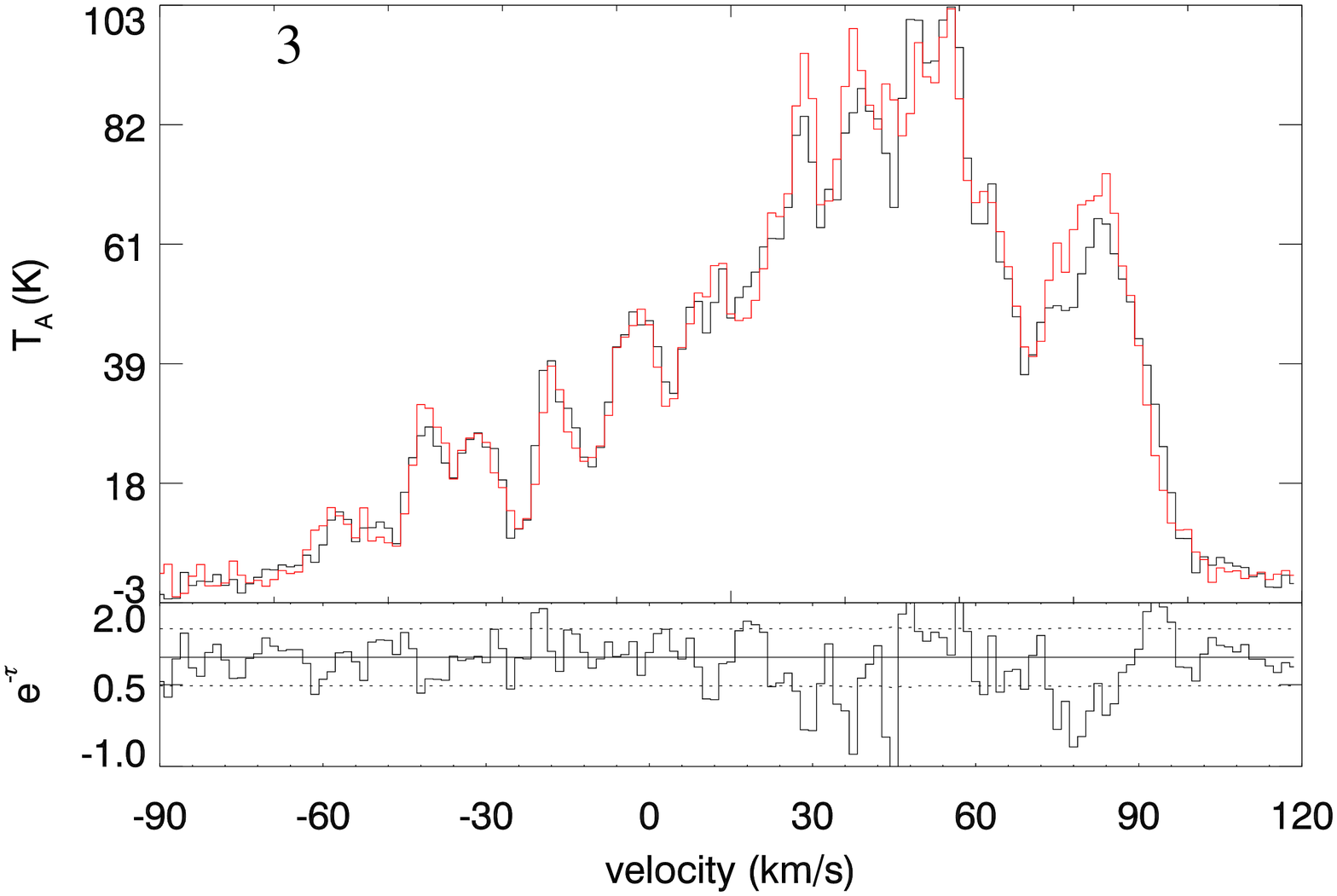}
\includegraphics[angle=0, width=0.45\textwidth]{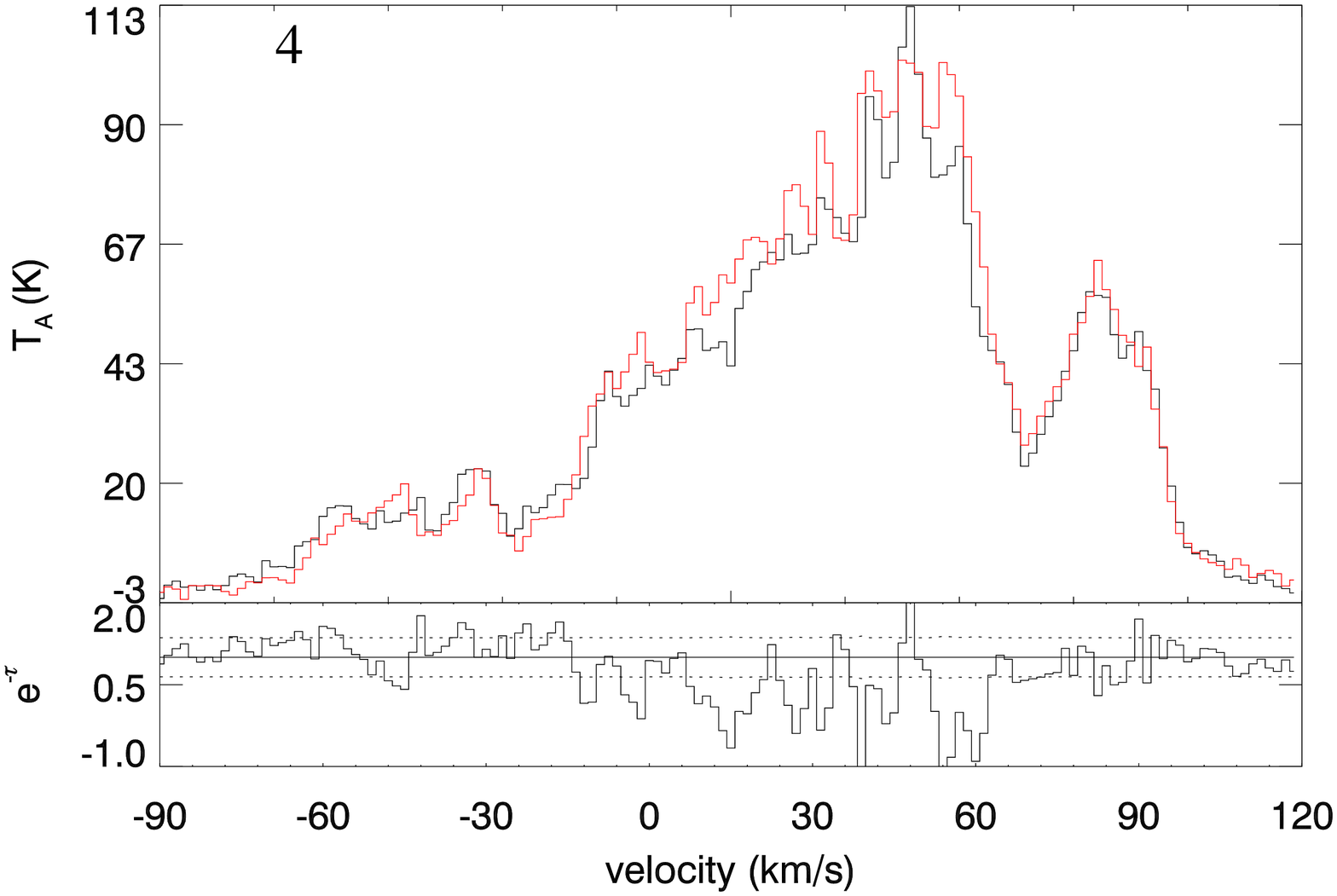}
\caption{
HI spectra (upper panels; red: background; black: source) and HI absorption spectra (lower panels) in three regions toward the SNR region.
The numbers in the left-top corner of each panel represent the regions 1, 2, 3, and 4 labeled in Figure~\ref{fig:rgb}.
The horizontal dotted lines in the lower panels indicate the $\pm 3\sigma$ uncertainty level of the HI absorption spectra.
}
\label{fig:HI}
\end{figure}

The systemic velocity of the eastern HIIR has been 
well measured as $\sim +56~\km\ps$ according to the RRLs \citep{lockman89} and CO observations \citep[][and also this study]{paron10}.
Assuming a flat Galactic rotation curve and adopting $R_0=8.34\, {\rm kpc}$ and $V_0=240\km\ps $ \citep{reid14},
this velocity corresponds to both a near distance $3.4\pm 0.4$~kpc 
and a far distance $10.2\pm 0.5$~kpc \citep{wenger18,reid14}, while the latter distance can be ruled out by HI absorption analysis.
We follow the methods of \cite{tian08} and plot the HI absorption spectra, where the HI and continuum data are taken from the THOR and VLA data.
The HI absorption is calculated as
$e^{\tau}=1-(T_{\rm off}^{\rm HI}-T_{\rm on}^{\rm HI})/(T_{\rm s}^{c}-T_{\rm bg}^{c})$, where $T_{\rm on}^{\rm HI}$ and $T_{\rm off}^{\rm HI}$ are HI temperatures at the source and background regions, and $T_{\rm s}^{c}$ and $ T_{\rm bg}^{c}$ are the radio continuum brightness temperatures at the source and background regions, respectively.
For the HIIR, the HI absorption spectrum is obtained from the region with label `4' in Figure~\ref{fig:rgb} and is displayed in the fourth panel of Figure~\ref{fig:HI}, showing that the maximum absorption velocity is $\sim+60\km\ps$.
This absorption feature is consistent with the results in \citet{zhu13} and supports the near distance.

SNR \snr\ was presumed to be associated with its eastern (in the equatorial coordinate system) HII region, as an MC revealed from the \thCO\ emission is projected at the eastern border of the SNR \citep{paron10}.
Our \twCO\ analysis has instead shown a strong morphological agreement between the SNR's western boundary and a warm molecular filament at $\VLSR\sim +50~\km\ps$.
The velocity corresponds to both a near distance $3.0\pm 0.4$~kpc and a far distance $10.5\pm 0.4$~kpc, which will below be discriminated with an HI absorption analysis. 

We inspect HI absorption toward the SNR, which has not been examined in previous literature, so as to put an additional constraint on the distance to it.
the HI and continuum data are taken from the THOR VLA data.
The HI absorption is calculated as
$e^{\tau}=1-(T_{\rm off}^{\rm HI}-T_{\rm on}^{\rm HI})/(T_{\rm s}^{c}-T_{\rm bg}^{c})$, where $T_{\rm on}^{\rm HI}$ and $T_{\rm off}^{\rm HI}$ are HI temperatures at the source and background regions, and $T_{\rm s}^{c}$ and $ T_{\rm bg}^{c}$ are the radio continuum brightness temperatures at the source and background regions, respectively.
The selected source and background regions are displayed in Figure~\ref{fig:rgb} with labels `1', `2', and `3'.
Figure~\ref{fig:HI} shows significant HI absorption in all of the three regions of the SNR at $V_{\rm LSR}\sim +80$~\kms, which corresponds to a distance of $4.7\pm0.7$~kpc or $8.9\pm0.8$~kpc.
Therefore, the SNR distance should be larger than $4.7\pm0.7$~kpc so as to explain the existence of the HI absorption at $\sim +80~\km\ps$.
Therefore, the SNR is located at the far distance $10.5\pm0.4\kpc$,
behind the tangent point toward the same line-of-sight (at $\sim 6.8\kpc$). 
Thus, SNR \snr\ and HIIR G35.6$-0.5$ are irrelevant to each other, given the different kinematic distances to them ($\sim 10.5\kpc$ and $3.4\kpc$, respectively).

By taking the two distances, the masses/densities of the MCs associated with SNR G35.6$-$0.5 and HIIR G35.6$-$0.5 are estimated to be about $9\times 10^3\,M_{\odot}$/$70\pcc$ and $1.3\times 10^3\,M_{\odot}$/$290\pcc$, respectively.

\subsection{Origin of gamma-rays}

\subsubsection{SNR \snr}
As shown in Figure~\ref{fig:tsmap_ab}c, SrcA is located within the shell of SNR \snr\ and is projectively close to two pulsars PSR J1857+0212 and PSR J1857+0210.
We first discuss the possibility of pulsar's origin.
According to the dispersion measures, PSR J1857+0212 and PSR J1857+0210 are at distances 7.98 kpc \citep{han2006} and 15.4 kpc \citep{morris2002}, respectively.
But the distances are revised to 6.0 kpc and 7.3 kpc in the ATNF pulsar catalog \citep{ATNF}, respectively, based on the new electron-density model developed by \citet{ymw17}.
Taking 6.0 kpc (the smaller one) for example, SrcA would have a luminosity of $\sim2\times10^{35}d_{\rm 6kpc}^{-2}\ {\rm erg\ s^{-1}}$ in the energy range 0.2--500 GeV, where $d_{\rm 6kpc}$is distance in unit of 6 kpc, which is significantly higher than the spin-down luminosity of $2.2\times 10^{34}\ {\rm erg\ s^{-1}}$ for PSR J1857+0212 \citep{han2006,ATNF} and $2.2\times 10^{33}\ {\rm erg\ s^{-1}}$ for PSR J1857+0212 \citep{morris2002,ATNF}.
Thus the pulsar's origin for SrcA can be ruled out unless the true distance is smaller than 2 kpc for PSR J1857+0212 and 0.7 kpc for PSR J1857+0210.

SNR \snr\ may be responsible for the \gray\ emission of SrcA, and we provide estimates for the emission in cases of hadronic and leptonic interaction separately.
We simply assume that the particles accelerated in the SNR have a power-law form with a high-energy cutoff:
\begin{equation}
    dN_i/dE = A_{i}(E/E_0)^{-\alpha_i} \mathrm{exp}(-E/E_{i,c})
\end{equation}
where, $i=e,p$, $\alpha_i$ and $E_{i,c}$ are the power-law index and high-energy cutoff, respectively. The normalization $A_i$ is determined by the total energy ($W_i$) in particles with energy above 1 GeV.
Since the SNR is very likely to be associated with the $\sim+50\km\ps$ MC at a distance of $\sim10.5$~kpc as revealed above, the \gray\ emission arising from the SNR-accelerated particles' hadronic interaction is first considered.
In such a hadronic scenario, SrcA's spectrum can be well fit (as shown in the left panel of Figure 7) and we obtain $\alpha_p\approx3.0$ and $n_0W_p\approx8\times10^{51}\ {\rm erg\ cm^{-3}}$, where $n_0$ is the number density of the target gas for proton-proton hadronic interaction.
The cutoff energy can not be constrained by the current data and is fixed as 3 PeV in our calculation.
With the mass of $9\times 10^3\,M_{\odot}$ and an average atomic hydrogen density $n_0\sim 140\ {\rm cm^{-3}}$ for the $\sim+50\km\ps$ filamentary molecular gas, the energy budget in protons $W_p$ is about $6\times10^{49}$ erg, which is acceptable and reasonable for the SNR scenario.
The somewhat large index may be attributed to the strong ion-neutral collisions \citep{Malkov2011} or the escaped process \citep[e.g.,][]{Aharonian1996,Li2012.9SNRs}.
In addition, the luminosity of SrcA in 1--100 GeV is $2\times 10^{36}\ {\rm erg\ s^{-1}}$, which is consistent with the known $\gamma$-ray-bright interacting SNRs \citep{Kes41.Fermi.Liu2015,acero16}.
Thus the hadronic scenario in which the energetic protons are from SNR \snr\ is a plausible explanation for SrcA.

\begin{figure}[h!]
\epsscale{0.55}
\plotone{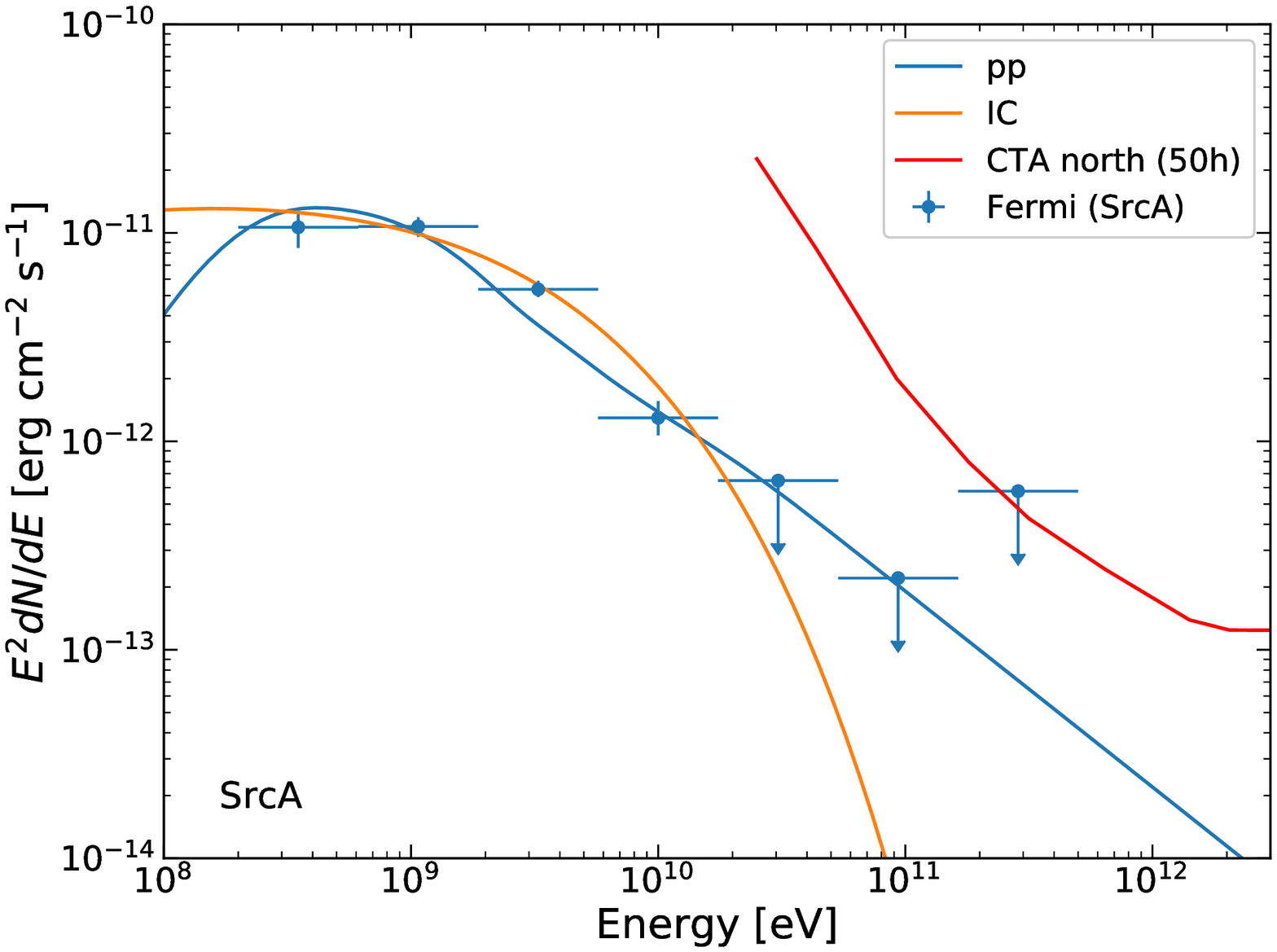}
\plotone{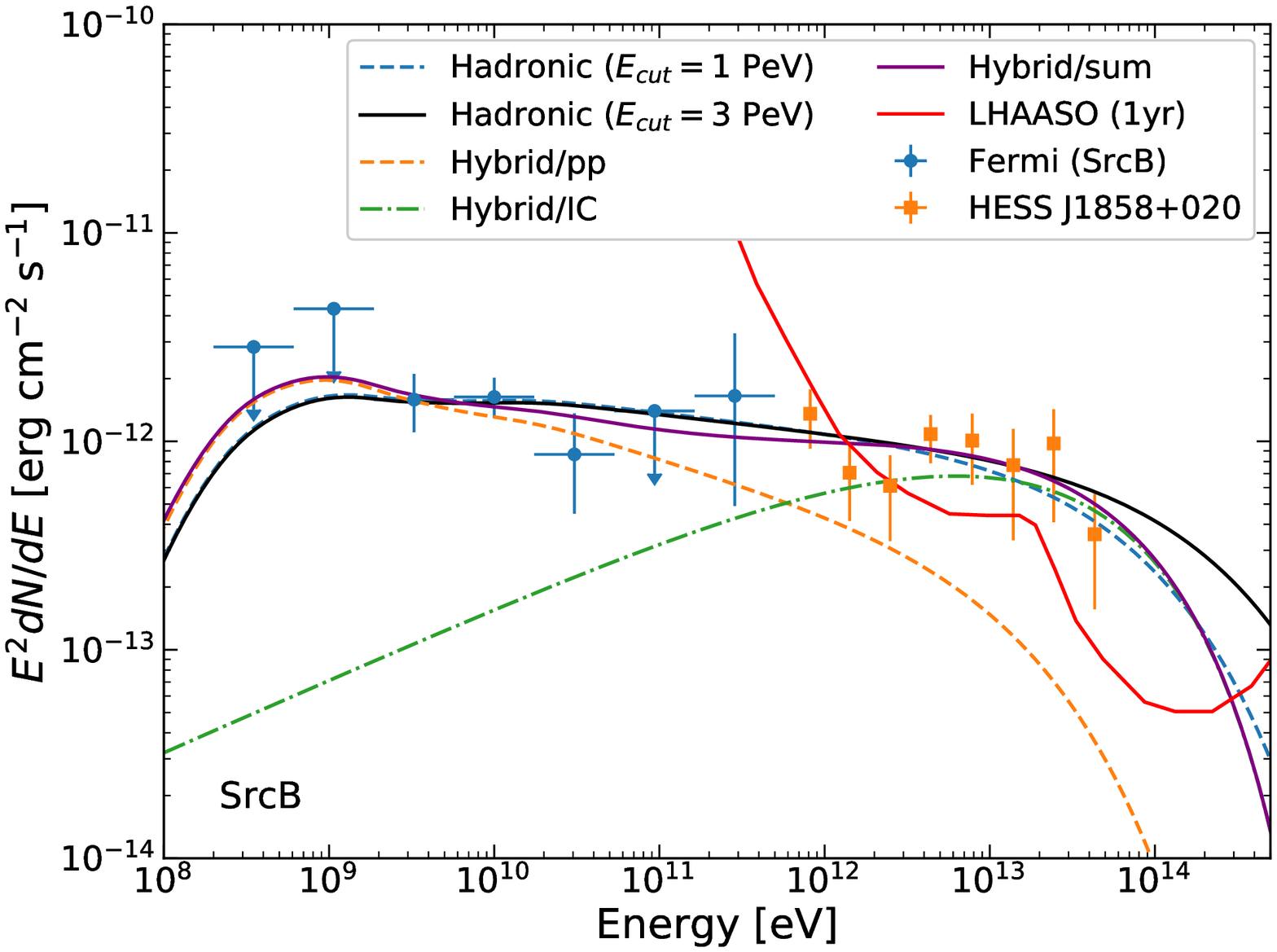}
\caption{
SED of SrcA (SNR {\snr}) and SrcB (HII region G35.6$-$0.5 and HESS J1858+020). The TeV data of HESS J1858+020 are taken from \citet{aharonian08}. The sensitivity of CTA north \citep[50h;][]{cta2019.book} and LHAASO \citep[1yr;][]{diSciascio2016} are also displayed in the left and right panel, respectively.
}
\label{fig:sed}
\end{figure}

For the leptonic case in which \grays\ are produced via the inverse Compton (IC) process,
due to the lack of constraint on the index by the GeV data, we fix $\alpha_e=2.0$ based on the radio index $\alpha_r=0.47$ \citep{green09}.
To explain the data (see the model curve in orange in the left panel of Figure~\ref{fig:sed}), $E_{e,c}\approx80$ GeV and $W_e\approx7\times10^{50}$ erg are obtained when the seed photons include
the IR emission with a temperature of 35 K and an energy density of 0.6 ${\rm eV\ cm^{-3}}$ estimated from the interstellar radiation field (ISRF) model \citep{Porter2006,shibata11} and the cosmic microwave background.
Considering the large amount of energy in electrons which is comparable to the canonical supernova explosion energy, the leptonic process seems not proper to explain the \gray\ emission of SrcA unless there is unusually strong IR emission.

\subsubsection{HIIR G35.6$-$0.5 and HESS J1858+020}
As shown in Figure~\ref{fig:tsmap_ab}d, the GeV source SrcB is spatially coincident with the unidentified TeV source HESS J1858+020.
Moreover, the GeV spectrum of SrcB can be well connected with the TeV spectrum of HESS J1858+020 by a simple power-law function with an index of $\sim$2.2 (see the right panel in Figure~\ref{fig:sed}).
Thus, SrcB is very likely the GeV counterpart of TeV source HESS J1858+020.
Meanwhile, both sources are spatially coincident with the MCs at a velocity around the $\sim$ +53--+57 \kms which have a mass of $1.3\times10^3\,M_{\odot}$ and an average atomic hydrogen number density of $600\ {\rm cm^{-3}}$.
As suggested in \citet{paron10} and this work (\S3.2), this molecular arc is likely associated with HIIR G35.6$-$0.5.
Both the association with MCs and the soft GeV-TeV spectra (index
$>$2) suggest the GeV-TeV \gray\ emissions likely have a hadronic origin.
One possible scenario for SrcB and HESS J1858+020 is that the energetic protons accelerated in HIIR G35.6$-$0.5 bombard the MC to produce the hadronic \gray\ emission.
By adopting a distance of 3.4 kpc, $\alpha_p\approx2.15$ and $W_p\approx 1.7\times 10^{47}$~erg are obtained.
To explain the TeV data which look without an obvious cutoff, the cutoff energy of protons $E_{p,c}$ should reach the PeV energy (see the right panel of Figure~\ref{fig:sed}).
This implies that there may be a potential PeV accelerator in HIIR G35.6$-$0.5.
This can be tested by the LHAASO experiment in the future.
As one of the PeVatron candidates, the lower limit of the proton cutoff energy $\sim$30 TeV for HESS J1858+020 is obtained by only using the TeV data \citep{Spengler2020}.
In fact, the index and the cutoff energy are strongly degenerated.
We note that the index constrained in \citet{Spengler2020} for this source is obviously smaller than 2.0.
With the help of the GeV data, the index is fitted as $\sim$2.2 in this study, resulting in a larger cutoff energy.

HIIRs, varies from $\sim$0.1 pc to hundred pc in size, are generally related to the active star formations and are considered as particle accelerators.
It has been proposed that particles can be accelerated by the colliding winds of binaries \citep[e.g.,][]{Eichler1993} and/or the young massive clusters \citep{Aharonian2019} which are the ionising sources of HIIRs.
This is supported by the TeV \gray\ observations towards $\eta$ Car \citep{etaCar.HESS2020} for the colliding winds and the Cygnus region \citep{veritas2014.cygnus,argo2014.cygnus,asgamma2021.cygnus,hawc2021.cygnus} for the massive star clusters.
In HIIR G35.6$-$0.5, although there are not massive stars detected as yet, \citet{paron10} found six young stellar object (YSO) candidates toward the molecular clump at $\sim$+53 \kms\ (namely the southern clump in their paper) and concluded that there is an active star-formation region. Further observations found at least one evolved YSO (i.e., IRS1) is embedded in
the clump, although no the signature of outflows was confirmed \citep{paron11}.
By analyzing of the Chandra data, \citet{paredes14} found seven X-ray sources, and suggested that sources X1--4, almost  projectively distributing on the shell of the HIIR (see the right panel of Figure~\ref{fig:rgb}), might be embedded protostars and that source X5, closing to the ``southern" (in the Galactic coordinate system) clump, might be coincident with the star formation region.
Thus, it could not be excluded that there may be some non-detected massive stars embedded in the MC.

%

Alternatively, we also consider the lepto-hadronic hybrid case in which the GeV emissions are from the hadronic process and the TeV \grays\ are generated by the IC process.
We assume the electrons and protons have the same power-law index and high-energy cutoff.
For the seed photons in the IC process, the IR component of ISRF is also included and is estimated as 35 K and 0.6 ${\rm eV\ cm^{-3}}$ by using the similar method for SrcA above.
To explain the data (see the purple curve in the right panel of Figure~\ref{fig:sed}), $\alpha_e=\alpha_p\approx2.3$ and $E_{c,e}=E_{c,p}\approx200$ TeV are obtained.
Electrons will suffer the synchrotron radiation loss during the acceleration, giving a cooling-limited maximum energy $E_{max,cool}=35u_3/\sqrt{\eta_g B_1}$~TeV \citep[e.g.,][]{Zirakashvili2007,Ohira2012}, where $u_3$ is the shock velocity in unit of 1000 \kms, $\eta_g$ the gyrofactor, and $B_1$ the magnetic field strength in unit of 10 $\mu$G.
For the wind velocity $u_3=1$ and the Bohm limit $\eta_g = 1$, it requires $B<0.3\ \mu$G to boost the energy of electrons up to 200 TeV. This magnetic strength is as weak as an order of magnitude lower than the mean value for Galactic ISM, which means that it is hard to accelerate electrons to the 100 TeV band for the standard interstellar magnetic field.
Thus, the hybrid model seems rather unlikely according to the fitted results.

\section{Summary}
\label{sec:summ}
In this study, we focus on the SNR-HIIR complex including SNR G35.6$-$0.4 and HIIR G35.6$-$0.5, which partially overlaps with the unidentified TeV source HESS~J1858+020 with a hard spectrum.
We reanalyze CO-line, HI, and Fermi-LAT GeV \gray\ emission data of this region.
The main results are summarized as follows:
\begin{enumerate}
    \item Based on the Nobeyama data, we found that a molecular arc at $\sim$+56 \kms\ delineates the northern (in the equatorial coordinate system) shell of HIIR G35.6$-$0.5 and a molecular filament at $\sim$+50 \kms\ nicely follows the western boundary of SNR \snr. Such morphological agreements, together with the relatively high main-beam temperature and the asymmetric or broad CO line profiles, suggest that the two molecular structures are likely to be associated with the HIIR and the SNR, respectively.
    
    \item The HI absorption features suggest that the SNR is located behind the tangent point, at a distance 10.5\,kpc, and thus is not associated with the HIIR.
    
    \item Performing the analysis of 12.3-year {\sl Fermi}-LAT data, we found that there are two point sources (SrcA and SrcB) with significance of 27.1$\sigma$ and 6.7$\sigma$ in 0.2--500 GeV toward the SNR-HIIR complex, respectively. The two sources are spatially coincident with the SNR and the TeV source HESS J1858+020, respectively.
    
    \item For SrcA,
    leptonic processes for the SNR scenario and the PSR scenario can be ruled out according to the energy budget. In the SNR-MC association scenario, the hadronic process can explain the spectrum with reasonable physical parameters. 
    
    \item For SrcB, its GeV-band spectrum can be smoothly connected with the TeV-band spectrum of HESS J1858+020 by a simple power-law function with an index of $\sim$2.2. In combination with the HIIR-MC association, it favors the hadronic origin. To explain the data, the cutoff energy of protons is the order of PeV. This indicates that there may be a potential PeV proton accelerator in HIIR G35.6$-$0.5, which needs to be tested with ultra-high energy observation, e.g., with LHAASO.
\end{enumerate}

\bibliography{sample63}{}
\bibliographystyle{aasjournal}

\acknowledgements
This publication makes use of data from FUGIN, FOREST Unbiased Galactic plane Imaging survey with the Nobeyama 45-m telescope, a legacy project in the Nobeyama 45-m radio telescope.
X.Z. thanks Xin Zhou, Siming Liu, and Jian Li for helpful discussions.
F.X.Z. thanks J. M. Paredes for proving the GMRT data.
We thank the support of National Key R\&D Program of China under nos. 2018YFA0404204 and 2017YFA0402600 and NSFC grants under nos. U1931204, 11803011, 12173018, 12121003, 11773014, 11633007, 11851305, and 12103049.

\software{
APLpy \citep{aplpy2012, aplpy2019}, 
Astropy \citep{astropy2013, astropy2018}, 
Fermipy \citep{fermipy}, 
Naima \citep{naima} 
}

\end{document}